%% file: paper.tex
\documentclass[sigconf]{acmart}

\AtBeginDocument{%
  \providecommand\BibTeX{{%
    \normalfont B\kern-0.5em{\scshape i\kern-0.25em b}\kern-0.8em\TeX}}}

\copyrightyear{2024}
\acmYear{2024}
\setcopyright{rightsretained}
\acmConference[CHI '24]{Proceedings of the CHI Conference on Human Factors in Computing Systems}{May 11--16, 2024}{Honolulu, HI, USA}
\acmBooktitle{Proceedings of the CHI Conference on Human Factors in Computing Systems (CHI '24), May 11--16, 2024, Honolulu, HI, USA}
\acmDOI{10.1145/3613904.3642230}
\acmISBN{979-8-4007-0330-0/24/05}

\usepackage{multirow}
\usepackage{float}
\usepackage{graphicx}
\newcommand{\rev}[1]{\textcolor{black}{#1}}

\begin{document}

\title{GazePointAR: A Context-Aware Multimodal Voice Assistant for Pronoun Disambiguation in Wearable Augmented Reality}

%
\author{Jaewook Lee}
\affiliation{%
  \institution{University of Washington}
  \city{Seattle}
  \state{WA}
  \country{USA}
}

\author{Jun Wang}
\affiliation{%
  \institution{University of Washington}
  \city{Seattle}
  \state{WA}
  \country{USA}
}

\author{Elizabeth Brown}
\affiliation{%
  \institution{University of Washington}
  \city{Seattle}
  \state{WA}
  \country{USA}
}

\author{Liam Chu}
\affiliation{%
  \institution{University of Washington}
  \city{Seattle}
  \state{WA}
  \country{USA}
}

\author{Sebastian S. Rodriguez}
\affiliation{%
  \institution{University of Illinois at Urbana-Champaign}
  \city{Urbana}
  \state{IL}
  \country{USA}
}

\author{Jon E. Froehlich}
\affiliation{%
  \institution{University of Washington}
  \city{Seattle}
  \state{WA}
  \country{USA}
}

\renewcommand{\shortauthors}{Lee, et al.}

\begin{abstract}
Voice assistants (VAs) like Siri and Alexa are transforming human-computer interaction; however, they lack awareness of users' spatiotemporal context, resulting in limited performance and unnatural dialogue. We introduce \textit{GazePointAR}, a fully-functional context-aware VA for wearable augmented reality that leverages eye gaze, pointing gestures, and conversation history to disambiguate speech queries. With GazePointAR, users can ask ``\textit{what's over \underline{there}?}'' or ``\textit{how do I solve \underline{this} math problem?}'' simply by looking and/or pointing. We evaluated GazePointAR in a three-part lab study (\textit{N}=12): (1) comparing GazePointAR to two commercial systems, (2) examining GazePointAR's pronoun disambiguation across three tasks; (3) and an open-ended phase where participants could suggest and try their own context-sensitive queries. Participants appreciated the naturalness and human-like nature of pronoun-driven queries, although sometimes pronoun use was counter-intuitive. We then iterated on GazePointAR and conducted a first-person diary study examining how GazePointAR performs in-the-wild. We conclude by enumerating limitations and design considerations for future context-aware VAs.

\end{abstract}



\begin{teaserfigure}
  \includegraphics[width=\textwidth]{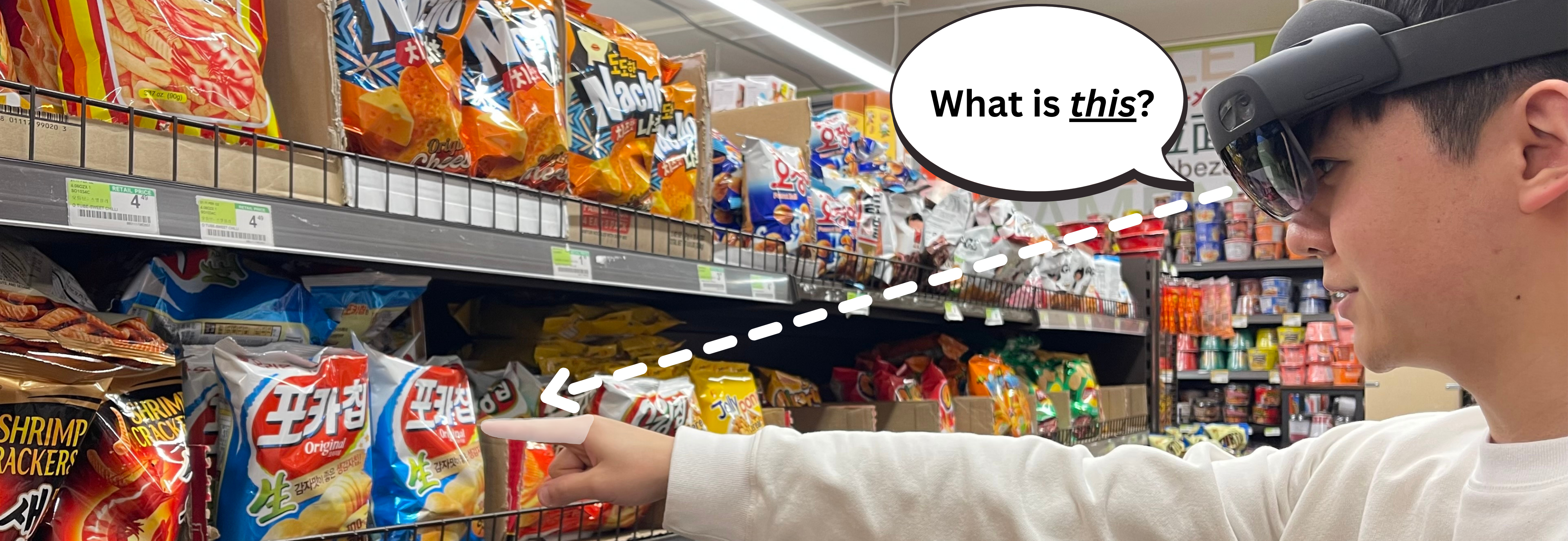}
  \vspace{-18px}
  \caption{An example interaction with GazePointAR. The user's query \textit{``What is \underline{this}?''} is automatically resolved by using real-time gaze tracking, pointing gesture recognition, and computer vision to replace ``\textit{this}'' with ``\textit{packaged item with text that says orion pocachip original},'' which is then sent to a large language model for processing and the response read by a text-to-speech engine.}
  \Description{An example interaction with GazePointAR. A user is gazing and pointing at a bag of chips with foreign language written on it and asking "What is this?".}
  \vspace{5px}
  \label{fig:banner}
\end{teaserfigure}

\maketitle

\input{sections/1_introduction}
\input{sections/2_related_works}
\input{sections/3_system_1}
\input{sections/4_study_1}
\input{sections/5_system_2}
\input{sections/6_study_2}
\input{sections/7_discussion}
\input{sections/8_conclusion}

\bibliographystyle{ACM-Reference-Format}
\bibliography{sample-base}

\input{sections/9_appendix.tex}









\end{document}

%% file: sections/1_introduction.tex
\section{Introduction}
Voice assistants (VAs) are transforming human-computer interaction. In a recent study of 2,000+ people~\cite{Olson2019}, 72\% of respondents indicated that they use VAs for tasks such as playing music, setting timers, controlling IoT devices, and managing shopping lists~\cite{Ammari2019, Bentley2018, Pradhan2018}. While widespread and useful, state-of-the-art VAs like Amazon Alexa, Google Assistant, and Apple Siri do not yet consider a user's spatiotemporal context, which can result in unnatural dialogue or unanswerable queries~\cite{Ammari2019}. For example, the query \textit{``What is \underline{that}?''} requires the VA to understand what ``\textit{that}'' refers to---a problem known as pronoun disambiguation~\cite{Corbett1983}. Despite their prominence in human speech~\cite{Diessel2020}, pronouns are not well supported by current VAs.

To resolve pronoun ambiguity, humans employ a variety of contextual clues, including eye gaze, pointing, and conversation history~\cite{Diessel2020}. For example, a person may physically gesture at an item in a store and ask ``\textit{How much is \underline{this}?}'' While straightforward for a human to resolve, current VAs are unable to answer this query precisely because they lack spatiotemporal context. Pronoun disambiguation and multimodal input have a rich history of research in HCI~\cite{Oviatt2000, Ruiz2010}—perhaps best marked by Bolt's visionary ``\textit{Put That There}'' system in 1980~\cite{Bolt1980} and beyond~\cite{Cohen1997, Koons1991, Zhai1999}. With recent advances in machine learning, speech recognition, and large language models (LLMs), new approaches are now possible. For example, emerging context-aware VA prototypes such as \textit{WorldGaze}~\cite{Mayer2020}, \textit{Nimble}~\cite{Romaniak2020}, and \textit{TouchVA}~\cite{Lee2021} examine how to use head gaze, pointing, and touch to resolve ambiguous queries. While promising and informative to our own work, these prototypes share similar limitations: they use \textit{Wizard-of-Oz} (WoZ) setups~\cite{Dahlback1993}, are accompanied by tightly-controlled lab studies \textit{vs.} open-ended queries, employ only one additional modality alongside speech, and are designed for smartphones rather than always-available head-worn displays.

In this paper, we introduce \textit{GazePointAR}, a context-aware VA for wearable augmented reality (AR), which uses eye gaze, pointing gestures, and conversation history to support pronoun disambiguation. If a user's spoken query contains a pronoun, we process the user's field-of-view using real-time computer vision, automatically extract objects and written text in the scene, and generate a new coherent query phrase that is sent to OpenAI's GPT-3~\cite{OpenAI} for processing. The response is then verbally read using speech synthesis. Pronouns are replaced using an empirically-tuned heuristic model that incorporates CV results based on gaze and pointing. For example, when asking ``\textit{How much is \underline{this}?}'' while looking at a bottle of mango juice (Figure~\ref{fig:system_overview}), GazePointAR extracts information such as object type, brand name, and flavor name to generate ``\textit{How much is \underline{a bottle with text that says Naked Mighty Mango 290 Calories}?}''.

To evaluate GazePointAR and explore the potential of context-aware VAs in wearable AR, we conducted two studies. First, we performed a three-part qualitative laboratory study with 12 participants to compare GazePointAR to two state-of-the-art query systems (\textit{i.e.,} Google Voice Assistant and Google Lens) (Part 1) and examine GazePointAR's usability and performance across various scenarios (Parts 2 \& 3). For example, participants searched for the price difference between two salt boxes (\textit{e.g.,} ``\textit{Can you compare the price between \underline{these} two?}''). In Part 3, we invited participants to brainstorm and try their own queries to further assess how context-aware VAs may be used in the future and how well GazePointAR currently supports such uses. Participants primarily used gaze to ask a diverse range of queries, from retrieving object information to foreign language translation, and were impressed by GazePointAR's ability to include their gaze to resolve queries. \rev{Participants also noted limitations, such as only capturing gaze data once after a query is spoken, the inability to handle queries with multiple pronouns, lack of AI explainability, and object recognition errors.}

Informed by these findings, we created a second GazePointAR prototype with improved object recognition and phrase generation techniques using prompt engineering, and conducted a follow-up first-person diary study~\cite{Desjardins2021}. Here, the first author used GazePointAR in their daily life for five days and recorded a written diary of usage, reflections, and observations of both successes and failures. In 20 hours of usage (4 hrs/day), the first author used GazePointAR across various contexts from cafes and restaurants to shopping malls and cinemas, and posed 48 queries, including recommendations for allergy-friendly menu items, ratings of movies, and cheaper alternatives to expensive clothing. Although the first author found GazePointAR to be more natural, instinctual, and robust against complex-to-describe objects in the real world than a traditional VA like Siri, they also encountered similar limitations as the study participants, such as static gaze data and limited object recognition capabilities, as well as privacy concerns with using a speech- and camera-based system in public.

In summary, our contributions include: (1) a fully-functional, context-aware VA for wearable AR that uses real-time computer vision and LLMs for pronoun disambiguation and more natural query dialogue; (2) findings from two user studies, including how users instinctively generate context-sensitive queries, how GazePointAR performs on queries from different scenarios, and limitations such as continuously tracking gaze information and AI explainability; and (3) a discussion on how to design future context-aware VAs that support any natural query a user poses spontaneously.

%% file: sections/2_related_works.tex
\section{Related Work} 
We provide background on pronoun usage in speech before enumerating relevant literature in multimodal interaction with a focus on voice assistants and augmented reality.

\subsection{Pronoun Usage in Speech}
Pronouns are frequently used in human speech, both in conversations between humans and in task-oriented dialogue systems---computational systems that complete tasks described in natural language. Leech \textit{et al.} ranked the frequency of 100 million spoken English words showing that pronouns, including demonstrative pronouns (\textit{e.g.,} ``\textit{this},'' ``\textit{that},'' ``\textit{these},'' ``\textit{those},'' ``\textit{here},'' and ``\textit{there}'') and third-person pronouns (\textit{e.g.,} ``\textit{it},'' ``\textit{he},'' ``\textit{him},'' ``\textit{she},'' ``\textit{her},'' ``\textit{they},'' and ``\textit{them}'') all ranked in the top 200~\cite{Leech2001}. As further evidence, Byron and Allen annotated a corpus of task-oriented dialogues and found that over one-third of 1,068 dialogue turns contained referential occurrences of pronouns ``\textit{it}'' and ``\textit{that}''~\cite{Byron1998}. Similarly, HCI studies have highlighted the importance of pronouns in human speech as they contribute to enhancing its naturalness and expressivity~\cite{Bolt1980, Lee2021, Khan2022} and that users desire to communicate to VAs using pronouns~\cite{Guindon1987}. To resolve pronoun ambiguity, humans rely on multimodality such as looking at or pointing at referents while speaking and conversational context~\cite{Diessel2020}. In our work, we investigate real-time gestures, eye gaze, and conversation history to enable pronoun disambiguation in human-VA interaction.

\subsection{Multimodal Interaction}
The HCI community has long been interested in multimodal interaction, highlighting various benefits such as improved naturalness, robustness, and expressiveness compared with unimodal interaction techniques~\cite{Oviatt2000, Ruiz2010}. For instance, researchers explored gaze as a multimodal input technique in mobile devices to address shortcomings of touch, such as slow interaction speed, limited reach on large screens, and impreciseness on small screens~\cite{Drewes2007, Mardanbegi2011, Pfeuffer2014, Esteves2015, Kong2021}. Additionally, gestures and speech have often been combined with gaze to improve the accuracy of gaze-alone systems~\cite{Chatterjee2015, Miniotas2006}. In our work, we rely both on \textit{gaze} and, if identified in the visual frame, \textit{pointing gestures} to resolve speech ambiguities. Many consumer products now support multiple modes of input, which allow users to interact using both touch and speech. Although the field of multimodal input is vast~\cite{Oviatt2000, Ruiz2010}, for the purposes of this paper, we focus primarily on its use in voice assistants and augmented reality.

\subsubsection{Multimodal Interaction with Voice Assistants}
The integration of speech with additional input modalities has long been a topic of interest in HCI. For example, Bolt's foundational ``\textit{Put That There}'' explored the use of speech and gestures as input~\cite{Bolt1980}. Further research has expanded on this idea by examining other input modalities, such as gaze pointing~\cite{Zhai1999}, pen and voice interaction~\cite{Cohen1997, Oviatt1992}, and merging speech, gestures, and eye gaze~\cite{Koons1991}. More recently, researchers have examined multimodal speech and gaze interactions in the context of hands-free communication between humans and vehicles~\cite{Aftab2019, Neßelrath2016, Roider2018}, as well as speech and gestures to support natural interactions with virtual objects in AR~\cite{Irawati2006, Piumsomboon2014, Liao2022}. Others have explored AR-based WoZ VA prototypes that support more natural dialogue between users and VAs by employing gaze~\cite{Mayer2020}, touch~\cite{Lee2021}, or gestures~\cite{Romaniak2020} alongside speech. The importance of multimodality in the design of voice user interfaces is widely acknowledged~\cite{Abdolrahmani2021, Diessel2020} because it enables flexible, expressive, natural, and contextual human-VA communication~\cite{Bolt1980, Guha2015, Khan2022}. Our work aims to contribute to this literature by implementing and evaluating a fully-functional multimodal VA with ambiguous speech support.

\subsubsection{Multimodal Interaction in Augmented Reality}
In AR specifically, multimodal interaction is frequently employed to improve object selection and manipulation, typically using hand gestures, gaze, and/or voice~\cite{Hertel2021, Williams2020}. \rev{For instance, both Olwal \textit{et al.} and Piumsomboon \textit{et al.} used speech as a supplement to gesture for improved object selection in AR~\cite{Olwal2003, Piumsomboon2014}}. Additionally, Kyt\"{o} \textit{et al.} used both head motion and eye gaze to increase the efficiency and accuracy of target selection in AR~\cite{Kyto2018}. \rev{Furthermore, Lystb\ae{}k \textit{et al.} used eye gaze to assist mid-air gestures with distant object selection in AR~\cite{Lystbaek2022}. Lastly, Liao \textit{et al.} used gestures and speech to generate and interact with AR presentation augmentations~\cite{Liao2022}.} Similarly, GazePointAR employs hand gestures to support gaze with a goal of enhancing real-world object selection.

Most relevant to our work, recent research has explored multimodal interaction in AR for pronoun disambiguation. More specifically, when a multimodal VA receives an ambiguous query, such as ``\textit{When does \underline{this} store open?}'', AR is used to analyze various visual contexts, including objects, texts, gaze, and gestures. For instance, Mayer \textit{et al.} presented WorldGaze, a WoZ smartphone-based multimodal VA that leverages head gaze information to clarify ambiguous queries~\cite{Mayer2020}. Others have explored touch~\cite{Lee2021} and pointing gestures~\cite{Romaniak2020} to resolve ambiguity. Each modality has tradeoffs: head gaze is quick and hands-free but can be inaccurate~\cite{Mayer2020}, touch is accurate but slower and not hands-free~\cite{Lee2021}, and gestures fall in between the two modalities~\cite{Romaniak2020}. In this work, we employ a combination of gaze supported by pointing gestures to create a efficient, mostly hands-free, and accurate input modality for speech disambiguation. We evaluate this in a fully-functional VA for wearable AR in various contexts.

\subsubsection{\rev{Other Uses of Gaze, Pointing, and Speech in Wearable AR}}
\rev{We conclude by highlighting recent studies that, while not employing gaze, pointing gestures, and speech as multimodal interaction techniques, present novel applications for each of these input sources in wearable AR. For instance, researchers have used eye gaze to design AR interfaces that adaptively control the display of information based on context, including its timing, placement, and volume~\cite{Pfeuffer2021, Lindlbauer2019, Lu2020, Rivu2020}. Additionally, hand gestures are often classified using machine learning to enable more natural object and UI manipulation~\cite{Pei2022, Saquib2019, Yan2018}. Furthermore, wearable AR glasses have been used to caption, translate, and augment speech in a non-intrusive way~\cite{Olwal2020, Jain2018a, Jain2018b, Liu2023, Rivu2020, Findlater2019, Jain2015, Miller2017, Schipper2017, Peng2018}. GazePointAR, while multimodal, is influenced by this prior work in wearable AR for enhanced interaction and context.}

%% file: sections/3_system_1.tex
\section{GazePointAR Prototype 1}
To advance the naturalness and economy of expression in how humans interact with VAs, we designed and built GazePointAR---a fully-functional context-aware VA for AR glasses that uses eye gaze, pointing gestures, and conversation history to support pronoun disambiguation. Below, we describe GazePointAR's design and implementation, starting with a taxonomy of pronoun usage drawn from linguistics literature.

\subsection{Taxonomy of Pronoun Use and Resolution}
To design GazePointAR, we first examined commonly-spoken pronouns in human speech and referent resolution strategies. We analyzed Leech \textit{et al.}'s ranked frequency list of 100 million spoken English words~\cite{Leech2001} and filtered to pronouns spoken at least 500 times per one million words. From this process, we extracted thirteen pronouns across three distinct groups of pronouns, all of which GazePointAR supports: nominal demonstrative pronouns: ``\textit{this},'' ``\textit{that},'' ``\textit{these},'' and ``\textit{those}'', adverbial demonstrative pronouns: ``\textit{here}'' and ``\textit{there}'', and third person pronouns: ``\textit{it},'' ``\textit{he},'' ``\textit{him},'' ``\textit{she},'' ``\textit{her},'' ``\textit{they},'' and ``\textit{them}''.

Demonstrative pronouns are used to point to specific people or things and can be further broken down into \textit{nominal} and \textit{adverbial}~\cite{Dixon2003}. In human conversations, gaze and/or pointing gestures are often used for referent disambiguation~\cite{Diessel2020}. While demonstrative pronouns such as ``\textit{this}'' and ``\textit{that},'' ``\textit{these}'' and ``\textit{those},'' and ``\textit{here}'' and ``\textit{there}'' seem similar, humans naturally employ one based on relative distance from the speaker to the referent~\cite{Diessel2020}. For example, a person may ask ``\textit{How much is \underline{this}?}'' when referring to a nearby object and ``\textit{How much is \underline{that}?}'' if the object is further away.

For third-person pronouns, ``\textit{it}'' may function as an \textit{anaphoric}, which refers to a word used previously in a phrase such as ``\textit{I have a bicycle. \underline{It} is red.}''; \textit{pleonastic}, which is the use of more words than needed to express meaning either unintentionally or for emphasis such as ``\textit{kick \underline{it} with your feet.}; or as an \textit{event reference} such as ``\textit{He lost his job. \underline{It} came as a total surprise.}''~\cite{Loiciga2017}. When resolving the anaphoric or pleonastic ``\textit{it},'' humans need prior conversation history, while for event reference, ``\textit{it}'' can be used interchangeably with ``\textit{this}'' or ``\textit{that}''~\cite{Guillou2016, Loiciga2017}. For other third person pronouns, humans often refer to entities such as other people or animals with ``\textit{he}'' or ``\textit{her}'', for example, but these pronouns must be used cautiously, as they can introduce gender bias~\cite{Cao2021}. 

Grounded in this analysis, we designed a taxonomy of frequently-spoken pronouns and how ambiguity from each pronoun can be resolved. When implementing GazePointAR, we adhered closely to this taxonomy, enabling our system to handle all thirteen pronouns and determine their referents based on gaze, pointing gesture, and conversation history.


\subsection{System Implementation}
We designed and implemented GazePointAR for the Microsoft HoloLens 2 with Unity 2021.3.16f1\footnote{\url{https://unity.com}} and Mixed Reality Toolkit (\textit{MRTK}) 2.8.2\footnote{\url{https://learn.microsoft.com/en-us/windows/mixed-reality/mrtk-unity/mrtk2}}. While our overarching vision is to develop an always available context-aware VA for lightweight AR displays, the HoloLens 2---despite its bulk---allowed us to rapidly prototype an implementation. 

\begin{figure*}[h]
  \centering
  \includegraphics[width=\linewidth]{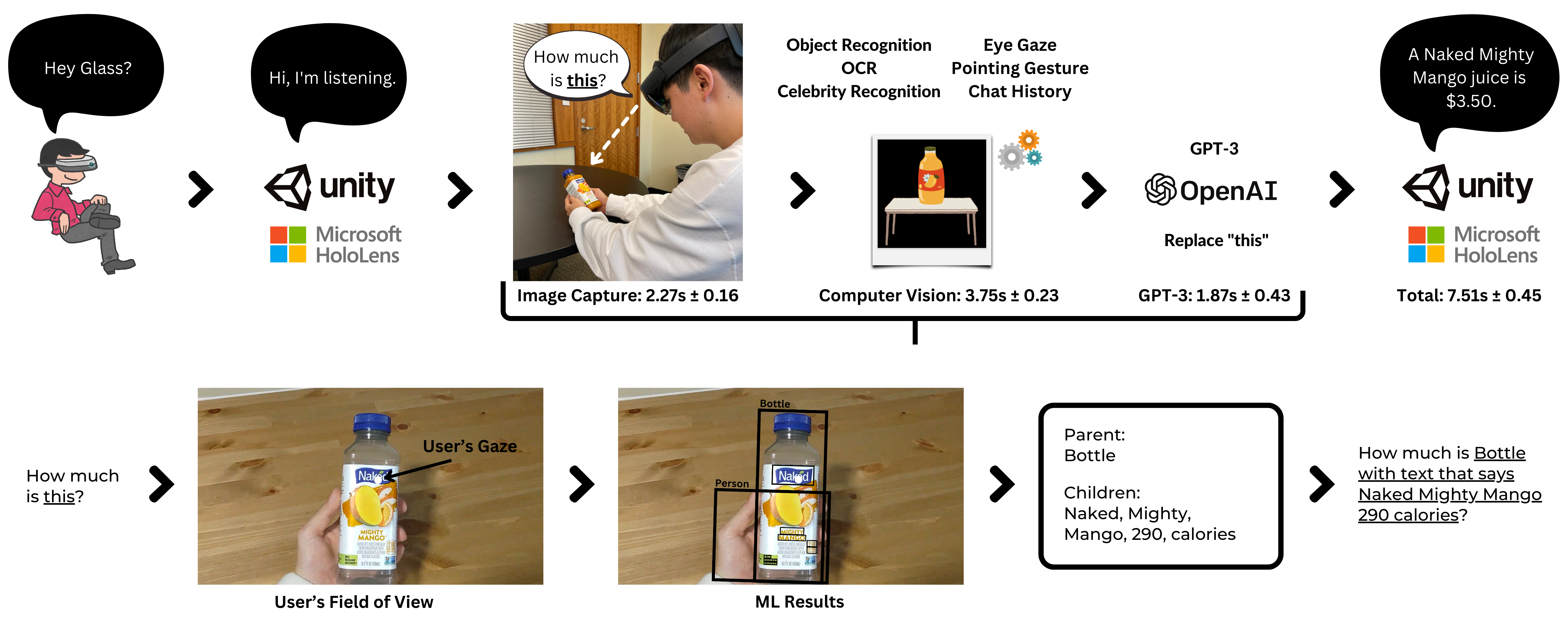}
  \caption{System overview and implementation details of GazePointAR}
  \Description{A system overview and implementation details diagram of GazePointAR. The user says "Hey Glass", to which the HoloLens response, "Hi, I'm listening." The user makes a spoken query, which in this figure is "How much is this?" GazePointAR captures an image, record user input data, and perform machine learning. These results are combined using a heuristric and the modified query is sent to GPT-3. The final response is read aloud, which in this figure is "A Naked Mighty Mango juice is \$3.50".}
  \label{fig:system_overview}
\end{figure*}

We designed GazePointAR to resemble the user experience of a commercial VA such as Apple Siri or Amazon Alexa. GazePointAR waits for a user to say \textit{``Hey Glass''} and make a verbal query. If the user's query contains one of thirteen pronouns in our taxonomy, it analyzes the user's field-of-view using various machine learning (ML) solutions, constructs a coherent phrase to describe the user's referent, replaces the pronoun with its referent, and sends the modified query to a large language model (OpenAI's GPT-3~\cite{OpenAI}). The query response is vocalized to the user using a text-to-speech engine within 10 seconds. See the system diagram in Figure~\ref{fig:system_overview}. We expand on key components below. As a rough examination of system response time, we asked the query ``\textit{How much is \underline{this}?}'' while gazing at a bottle of mango juice (a tutorial task) ten times. GazePointAR responded in $7.51\pm0.45$ seconds. We include sub-component performance times from this same procedure below.

\textbf{Activating GazePointAR.} To activate GazePointAR, the user states ``\textit{Hey Glass.}'' For this, we implemented a continuously-running background process checking for the trigger phrase. Upon recognition, GazePointAR replies, ``\textit{Hi, I'm listening.}'' and waits for a spoken query. After the query, GazePointAR performs a substring search to check for pronouns from our taxonomy.

\textbf{Capturing and analyzing the user’s field-of-view.} If the query contains a pronoun, GazePointAR prompts the HoloLens to take a 1080p photo of the user's field-of-view. For user and bystander privacy, the captured image is stored temporarily and deleted once a query response is received. \rev{This process takes $2.27\pm0.16$ seconds to complete.}

Once the user's field-of-view is captured, we begin analyzing the image for objects, texts, and faces. We send the captured image to three ML models through asynchronous POST requests to minimize runtime: \textit{Google Cloud Vision}'s (1) \textit{Object Localization} and (2) \textit{Optical Character Recognition} (OCR) models~\cite{GoogleCloudVision}, as well as (3) \textit{Amazon Rekognition's Celebrity Recognition} model~\cite{AmazonRekognition}. \rev{This process takes $3.37\pm0.23$ seconds to complete.}

After receiving JSON responses from the ML services, GazePointAR identifies hierarchical relationships between the detected objects, faces, and texts. We treat the object detection and celebrity recognition results as the parent layer. The child layer, comprised of OCR results, is connected to parent bounding boxes that have at least 70\% pixel overlap (a threshold tuned empirically). Each parent can have up to five OCR results, ranked by bounding box size. This ensures that GazePointAR prioritizes important textual information, such as product and brand names, which tend to be larger in the user's field-of-view, while ignoring less important, smaller details like promotional blurbs. For example, as shown in Figure~\ref{fig:system_overview}, when a user asks ``\textit{How much is \underline{this}?}'' while holding a bottle of Naked Mighty Mango juice, possible parent layer objects include ``\textit{person}'' and ``\textit{bottle}'', with ``\textit{bottle}'' having child layer objects such as ``\textit{Naked}'', ``\textit{Mighty}'', ``\textit{Mango}'', ``\textit{290}'', and ``\textit{calories}''.

\textbf{Gaze tracking and gesture recognition.} To capture the user's eye gaze and pointing gesture, we customized MRTK's built-in gaze and pointer modules. For gaze, we designed a white sphere that follows the user's gaze from a fixed distance (\textit{i.e.,} 2 meters) and is overlaid in their field-of-view. This allows us to retrieve 3D gaze coordinate data and also provides visual feedback to the user about their system-inferred gaze. 

For pointing, we implemented a finger-pointing gesture to supplement the base palm-pointing gesture, since extending the arm and index finger is a more typical pointing gesture~\cite{Diessel2020}. Performing a pointing gesture creates a ray that extends away from the user's hand until a collision with an object in the physical world occurs. To achieve this, we integrated MRTK's spatial awareness into GazePointAR to detect collisions between user inputs and spatial meshes generated in real-time.

As the HoloLens captures an image, GazePointAR simultaneously logs the locations of both the user's gaze and pointing gesture. To convert 3D gaze and pointing gesture coordinates to their corresponding pixel locations on the captured image, we use projection.

\textbf{Query assembly and pronoun replacement.} 
Using the ML-generated results and pixel coordinates of gaze and pointing gesture, GazePointAR assembles a coherent phrase to replace the user-spoken pronoun. To accomplish this, we employ a state diagram, which encompasses the differences in pronouns in our taxonomy. 

If a pronoun is singular, GazePointAR computes whether any input coordinates fall within any parent (\textit{i.e.,} object recognition and celebrity recognition results) bounding boxes. If so, GazePointAR takes that parent object's child layer (\textit{i.e.,} OCR results) and creates the following phrase: ``\textit{[parent] with text that says [children]}''. Otherwise, GazePointAR takes the five nearest child layer texts from each input coordinate, computes a union, orders them by distance, and uses the five closest to build the following phrase: ``\textit{[OCR Result 1] [OCR Result 2] ... [OCR Result 5]}.'' For example, returning to the price of a Naked Mighty Mango juice in Figure~\ref{fig:system_overview}, a user is looking at the ``\textit{Bottle}'', meaning GazePointAR generates the phrase ``\textit{Bottle with text that says Naked Mighty Mango 290 calories}''.

If the pronoun is plural, GazePointAR expands the gaze and pointing gesture pixel coordinates into bounding boxes with width and height equivalent to half of the captured image's width and height. Then, GazePointAR computes whether any input bounding boxes have at least 70\% overlap with any parent bounding boxes. The rest of the procedure is the same as with singular pronouns.

\textbf{Answering the query.} GazePointAR assembles the final query by combining the user-spoken query, the ML-generated phrase, and text from the five most recent query-answer pairs. The final result is processed by OpenAI's GPT-3~\cite{OpenAI}\rev{, which takes $1.87\pm0.43$ seconds to complete}. The output is displayed as text and read aloud. If there are no ML results or GPT-3 cannot process the modified query, GazePointAR responds ``\textit{Sorry, I did not understand your question.}'' Users can ask follow-up questions or provide additional information appropriately.

%% file: sections/4_study_1.tex
\section{Study 1: Three-Part Lab Evaluation of GazePointAR}

To evaluate GazePointAR and explore the potential of context-aware VAs in wearable AR, we conducted two studies: (1) a laboratory study to compare GazePointAR to two state-of-the-art query systems and examine how participants generate and use their own context-sensitive queries; and (2) a first-person diary study using GazePointAR in the real world. We report on the first study below.

For the lab study, we sought to address three primary research questions: (1) How do users initially perceive and use a multimodal, context-aware VA for pronoun disambiguation? (2) How does performance compare to traditional VAs? (3) What types of queries do users want to perform with a context-aware VA, and how well does GazePointAR support these queries? As initial work, our primary aim was not to quantitatively examine GazePointAR's performance but rather to observe how participants reacted to and used a fully-functional, context-aware query system for AR glasses.

To address these questions, we conducted a three-part, within-subjects laboratory study with 12 participants. In Part 1, we asked participants to complete a common query task with GazePointAR as well as two state-of-the-art commercial systems: \textit{Google Voice Assistant} (\textit{voice input}) and \textit{Google Lens} (\textit{image+text input}). In Part 2, participants completed three additional context-dependent query tasks with GazePointAR, which were designed to highlight different aspects in our design space (\textit{e.g.,} pronoun use, gaze, gesture, and conversation history). Finally, in Part 3, participants brainstormed and tried their own context-sensitive queries.

\subsection{Participants}
We recruited 12 participants via mailing lists, social media, and snowball sampling. Participants were screened via a demographic questionnaire, which asked about prior experiences with VAs, AR, and AI chat systems. Given the reliance on gaze and speech in our study, we filtered participants who indicated visual or auditory disabilities, have a history of seizures or epilepsy, or are not fluent in English. All twelve participants indicated at least some previous experience with VAs, including \textit{Amazon Alexa}, \textit{Apple Siri}, \textit{Google Voice Assistant}, \textit{Microsoft Cortana}, and \textit{Samsung Bixby}. Most (9/12) had not previously used AR headsets or glasses—those that did (3/12) mentioned Google Glass, Microsoft HoloLens, and Meta Quest Pro. Finally, all participants indicated at least some familiarity with AI chat systems with six stating that they use them at least once a week (two participants marked never). Most commonly, participants mentioned ChatGPT (8/12) and customer support chatbots (2/12).

\subsection{Procedure}
The in-person laboratory study took place on a university campus and lasted 60 minutes. Instructions were presented orally with backing slides to improve comprehension. Consent and background forms were emailed in advance; written consent was taken in person. All sessions were video recorded for \textit{post hoc} analysis. Because we were interested in candid reactions, we did not tell participants that we created GazePointAR. 

\textbf{Tutorial.} After consenting, participants completed a short tutorial about each VA system: GazePointAR, Google VA, and Google Lens. The tutorial order was counterbalanced but the query task was the same: ``\textit{Your task is to find the price of this bottle of Naked Mighty Mango juice}'' (Figure~\ref{fig:system_overview}). During the tutorial, participants could ask questions of the study facilitator and, for GazePointAR, configure the AR headset fit and calibration. The study commenced once each participant was comfortable with all three VA systems.

\begin{figure*}[h]
  \centering
  \includegraphics[width=\linewidth]{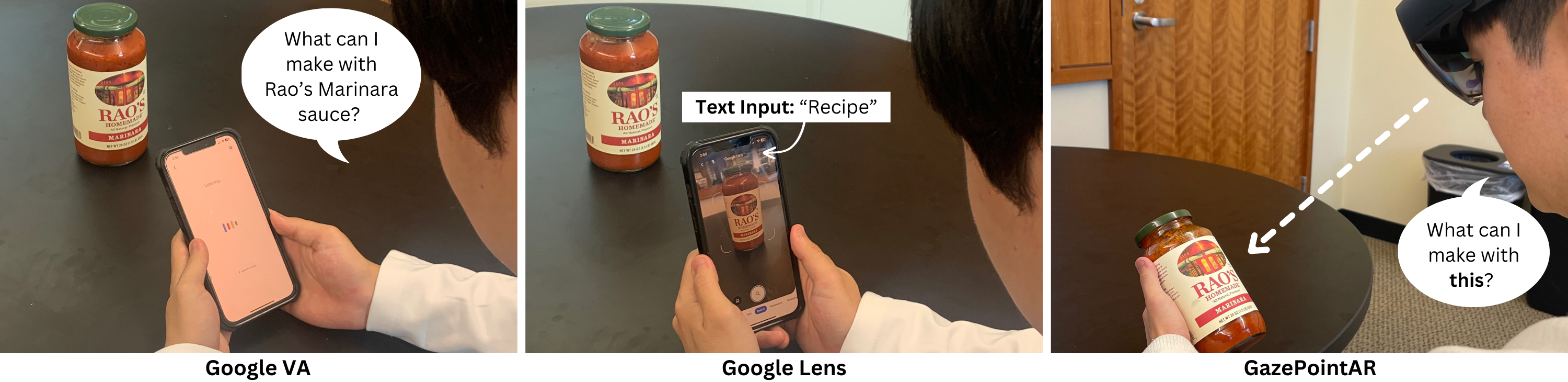}
  \caption{Cooking scenario and the three VAs used in Part 1 of the study.}
  \Description{Cooking scenario and the three VA systems used in Part 1 of the lab study. On the left, a participant is using Google Voice Assistant and asking "What can I make with Rao's Marinara sauce?" In the middle, a participant is using Google Lens and taking a photo of the jar of marinara sauce and typing in text input that reads "Recipe". On the right, a participant is using GazePointAR and asking "What can I make with this?" while gazing at a jar of marinara sauce.}
  \label{fig:part1}
\end{figure*}

\textbf{Part 1: Comparing VAs.} Part 1's goal was to examine how participants constructed queries for a common VA scenario: cooking. Specifically, we asked participants to ``\textit{find a marinara pasta recipe that uses this jar of Rao's Marinara sauce; the more specific, the better}'' (Figure~\ref{fig:part1}) using each of the VA systems---which were again counterbalanced. For each VA system, we encouraged participants to construct the query to best leverage the system's input modality (\textit{e.g.,} taking a picture for Google Lens, gazing or pointing for GazePointAR). The search task was deemed complete when the participant had found, from their perspective, a satisfactory recipe. After using each VA, participants filled out a \textit{System Usability Scale} (SUS)~\cite{Brooke1996, Peres2013} questionnaire and answered interview questions regarding their experience. At the end of Part 1, we asked participants to rank the three systems in terms of perceived intelligence, helpfulness, naturalness, and overall preference. We then asked follow-up questions to justify rankings.

\begin{figure*}[h]
  \centering
  \includegraphics[width=\linewidth]{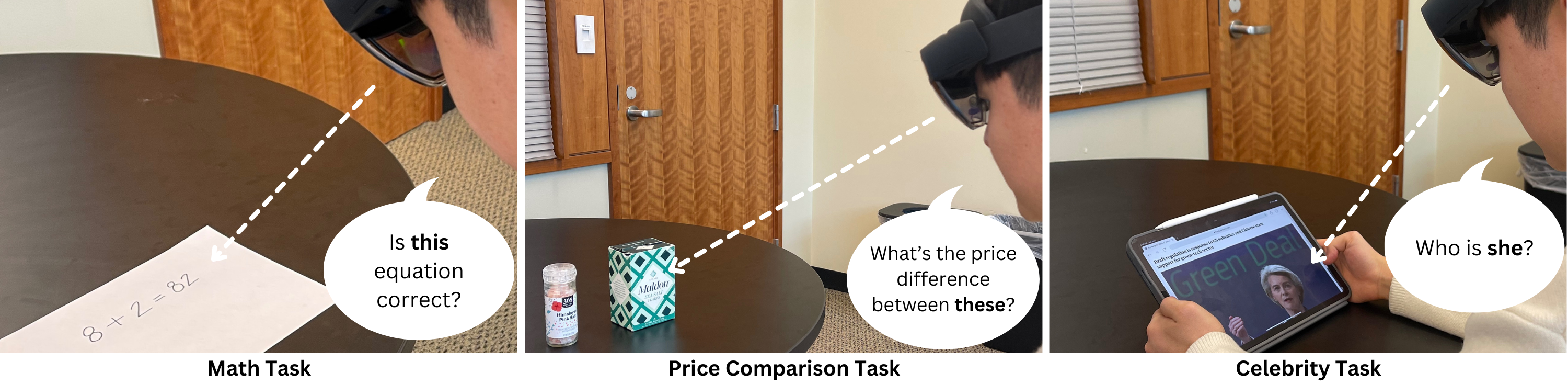}
  \caption{Usage scenarios in Part 2 of the study.}
  \Description{Three scenarios used in Part 2 of the lab study. On the left is a math task, and a participant is asking "Is this equation correct?" while gazing at a sheet of paper with an equation written on it. In the middle is a price comparison task, and a participant is asking "What's the price difference between these?" while gazing in between two salt boxes. On the right is a celebrity task, and a participant is asking "Who is she?" while gazing at a picture on an online article.}
  \label{fig:part2}
\end{figure*}

\textbf{Part 2: Context-sensitive Queries with GazePointAR}. While Part 1 examined differences in query behavior depending on modality and technology, Part 2 specifically focused on examining context-sensitive queries with GazePointAR. We asked participants to complete three tasks that, based on our own usage of GazePointAR, benefited from context-dependent queries and pronoun disambiguation: (1) \textit{Write a simple math equation on a sheet of paper and ask GazePointAR if it is mathematically accurate}; (2) \textit{Use GazePointAR to find the cost difference between two items}; (3) \textit{Use GazePointAR to find more information about a person in a magazine article} (Figure~\ref{fig:part2}). Again, at the end of Part 2, we asked participants to remark on their GazePointAR experiences and the additional search tasks.

\begin{figure*}[h]
  \centering
  \includegraphics[width=\linewidth]{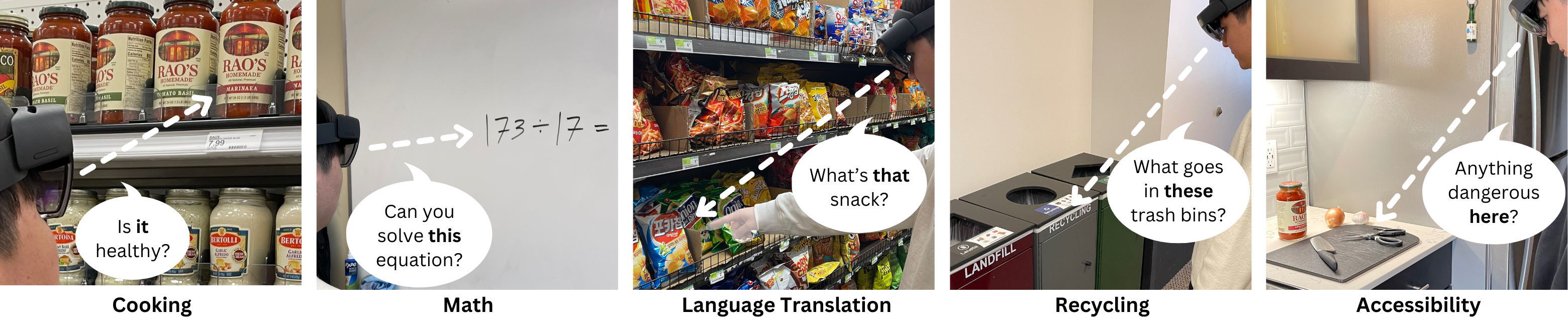}
  \caption{Design probes in Part 3 of the study. See supplementary materials for the videos.}
  \Description{Design probes used in Part 3 of the lab study. On the left is a cooking task, and a person is asking "Is it healthy?" while gazing at a jar of marinara sauce. On the middle left is a math task, and a person is asking "Can you solve this equation?" while looking at a white board with a math equation written on it. In the middle is a language translation task, and a person is asking "What is that snack?" while gazing and pointing at a bag of chips with foreign language written on it. On the middle right is a recycling task, and a person is asking "What goes in these trash bins?" while gazing at a group of trash bins. On the right is an accessibility task, and a person is asking "Anything dangerous here?" while looking ahead.}
  \label{fig:part3}
\end{figure*}

\textbf{Part 3: Design Probe and Co-design}. Finally, in Part 3, participants helped co-design the future of context-aware VA systems. Using a design probe method similar to Mauriello \textit{et al.}~\cite{Mauriello2015}, participants first watched five video clips of GazePointAR being used across diverse scenarios: cooking, math, language translation, recycling materials, and asking if there are dangerous items nearby (Figure~\ref{fig:part3}). After viewing and discussing the design probe videos, participants brainstormed and then actually attempted their own context-sensitive queries---a study task that is only possible with a fully-functional prototype like GazePointAR.

\subsection{Data and Analysis}
\begin{figure*}[h]
  \centering
  \includegraphics[width=0.8\linewidth]{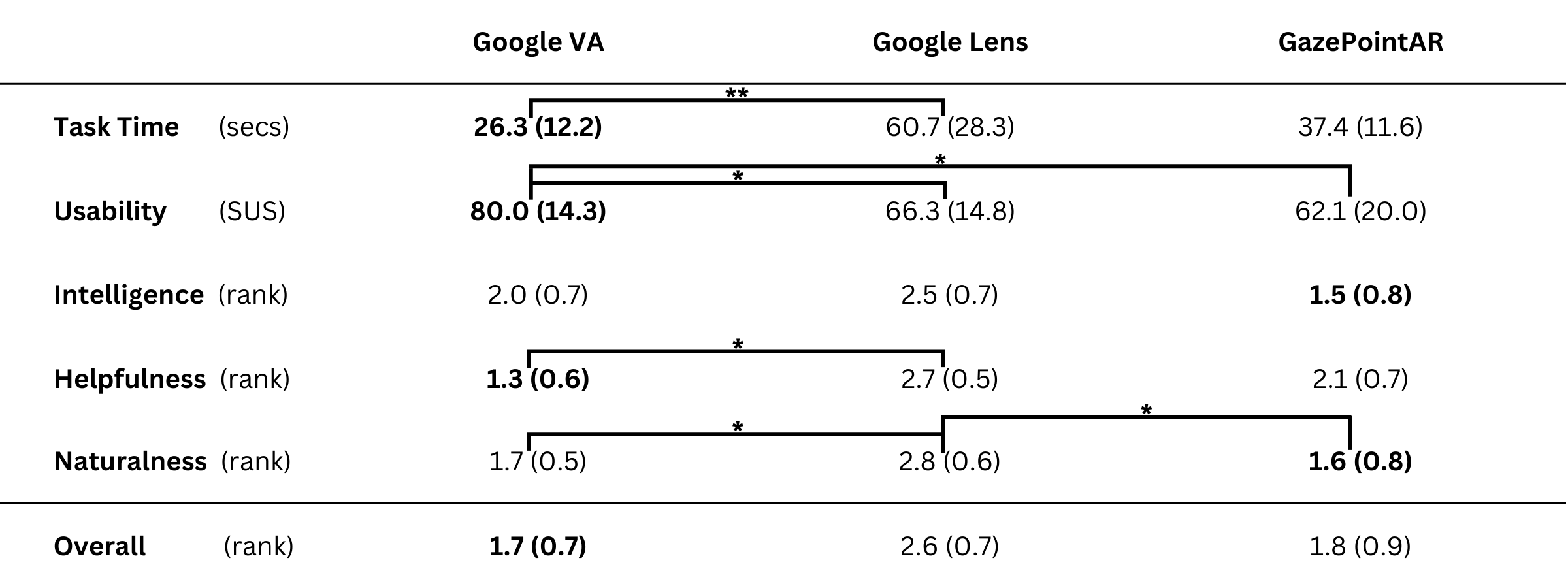}
  \caption{The mean and standard deviation of task time, usability, perceived intelligence, helpfulness, naturalness, and overall preference. Task Time is in seconds. Usability is 0-100; higher the better. Rankings are 1-3; lower is better. For statistical significance, one asterisk (*) is $p < 0.05$; two asterisks (**) is $p < 0.01$.}
  \Description{A graph showing average and standard deviation of task time, usability, intelligence, helpfulness, naturalness, and overall of Google VA, Google Lens, and GazePointAR. Two asterisks for Task Time Google VA > Google Lens. One asterisk for Usability Google VA > Google Lens, Google VA > GazePointAR. One asterisk for Helpfulness Google VA > Google Lens. One asterisk for Naturalness GazePointAR > Google Lens and Google VA > Google Lens.}
  \label{table:rankings}
\end{figure*}

We analyzed three sources of data: interview transcripts, observations from the user study sessions, and the post-task questionnaires. For the qualitative data, we used reflexive thematic coding~\cite{Braun2006, Braun2019}. The first author, who facilitated all user study sessions, created an initial codebook by reviewing study transcripts. The entire team then collaboratively iterated on the codebook while checking for bias and coverage. With a final codebook consisting of 34 codes, the first author coded participants' quotes, after which the team discussed the resulting themes. While this exploratory study focused on participants' reactions to GazePointAR, we also collected quantitative data from Part 1 to compare GazePointAR with existing systems. For SUS scores, we converted survey responses, which are on a scale of 0-40 when summed, to a range between 0-100\footnote{(((Q1 + Q3 + Q5 + Q7 + Q9) - 5) + (25 - (Q2 + Q4 + Q6 + Q8 + Q10))) * 2.5}~\cite{Brooke1996}. We then conducted a Friedman test as an omnibus test with an appropriate number of Wilcoxon signed-rank tests corrected with Holm’s sequential Bonferroni procedure for statistical significance. See Figure~\ref{table:rankings} for a summary of quantitative results.

\subsection{Findings} 
We report key findings, including how VA input modality influenced perceived performance and query formation, the queries participants generated using GazePointAR, and successes and failures of GazePointAR in various scenarios. We denote each participant as P\# (\emph{e.g.,} P1 for participant 1). Quotes have been lightly modified for concision and clarity.

\subsubsection{Part 1: Comparing VAs} \hfill

\noindent In Part 1, participants completed an open-ended query task to find a recipe for a specific marinara sauce with the three different VA systems (Figure~\ref{fig:part1}). We first provide overall reactions before analyzing query formations, perceived intelligence, naturalness, and helpfulness, task completion time, and usability. 

\sloppy\textbf{Overall.} Overall, participants preferred using Google VA (\textit{$mean_{rank}=$}1.7; \textit{SD=}0.7) and GazePointAR (1.8; \textit{SD=}0.9) over Google Lens (2.6; \textit{SD=}0.7)---lower is better, range is 1-3. For Google Lens, participants emphasized that while taking photos was familiar (3/12) and lessened the specificity of their queries compared to voice-only systems (3/12), manually capturing an image and supplying written text felt tedious (3/12) and unnatural (2/12). As P2 stated, ``\textit{I had to take a picture and then add more information... It's like an extra step, right? Is this necessary?}''. Similarly, P4 said, ``\textit{Google Lens is the most unnatural, because sometimes you have to type extra context, and I feel like that's just another hurdle.}''. Finally, the quality of Google Lens' responses influenced opinions: four participants were initially guided to a recipe for \textit{making} marinara sauce rather than \textit{using} Rao's Marinara sauce. Two participants mentioned losing confidence in Google Lens due to poor responses.

For Google VA, participants appreciated the straightforward (6/12), quick (4/12), and hands-free (2/12) nature of the system. Additionally, four participants emphasized that, compared to Google Lens and GazePointAR, it was easier to review query responses, visit different links, and decide on the best answer themselves. As P5 said, ``\textit{Google voice assistant displayed a typical Google search result [on the phone], which gives me a lot of options... clicking into them allows you to try until you find the recipe that you're satisfied with.}'' Half of the participants also mentioned the familiarity of Google VA and the results interface. For limitations, participants noted that Google VA requires queries to be highly specific (4/12), necessitates accurate pronunciation of complex words like ``\textit{Rao's}'' (4/12), and leads to longer queries, which are laborious to say (3/12). P12 aptly summarized theses issues by stating, ``\textit{You have to be more specific and have to say a lot more... I also think that a lot of people might mispronounce Rao's.}'' One participant (P3) felt strongly about voice-edit capabilities—as Google VA only allows query iteration through text but not voice-based editing.

Finally, for GazePointAR, participants felt that it was simpler (8/12) and faster (8/12) to interact with as well as more natural (7/12) and human-like (6/12) to speak to than Google VA and Google Lens. In part, this was because participants could reduce the specificity of their queries with GazePointAR's context-awareness features. As P10 said, ``\textit{When speaking to GazePointAR, I am giving it a voice input while also interacting with the product that I am talking about. Perceptually, this is the most natural way of speaking, which is why we do this when talking to other people as well.}''. Another said: ``\textit{When you're talking to someone, you point to or look at something and say 'what is \underline{this}?' They can see what you're pointing to or looking at, which is exactly what the headset is doing... I was also able to receive an answer quickly without having to look through web pages.}'' (P4). However, the most common criticism (8/12) was that GazePointAR provided only a single answer rather than an interactive, explorable list like a traditional search engine. Participants also requested more transparency from the system about their gaze and pointing gestures, the image GazePointAR took for scene processing, desired citations in the query response, and wanted queries to be editable.

\textbf{Query formations.} Beyond overall reactions, we also explored \textit{how} participants formed queries with the three systems. When examining query length, unsurprisingly, the two multimodal systems had shorter queries on average: Google Lens (\textit{avg=}1.3 words long; \textit{SD=}0.5) and GazePointAR (\textit{avg=}6.3; \textit{SD=}1.8) than Google VA (\textit{avg=}8.4; \textit{SD=}2.2). With Google Lens, all participants took a picture of the sauce jar then supplied additional text, including ``\textit{recipe}'' (9/12) and ``\textit{recipe using}'' (3/12). With GazePointAR, all participants used the pronoun ``\textit{this}'' along with gaze but did not use pointing. P2 reasoned that ``\textit{If you're pointing at something, you have to use your hand. This implies that you still have use of your hands during some tasks. Also, because the jar is so close, the system shouldn't need pointing to tell what I'm talking about.}'' Finally, with Google VA, all participants used proper nouns, including various formations of ``\textit{Rao's homemade Marinara sauce}''. Full queries are in Appendix ~\ref{table:queries1}.

\textbf{Perceived intelligence, helpfulness, and naturalness}. For perceived intelligence, participants ranked GazePointAR the highest with \textit{$mean_{rank}$=1.5} (\textit{SD=}0.8), followed by Google VA (2.0; \textit{SD=}0.7), then Google Lens (2.5; \textit{SD=}0.7). A majority of participants (8/12) reasoned that GazePointAR ``\textit{recognized things I am talking about just from my gaze and pointing}'' (P3), while for Google VA and Google Lens, ``\textit{instead of it figuring things out itself, I have to provide everything}'' (P12). For perceived helpfulness, participants ranked Google VA the highest with \textit{$mean_{rank}$=1.3} (\textit{SD=}0.6), followed by GazePointAR (2.1; \textit{SD=}0.7), then Google Lens last (2.7; \textit{SD=}0.5). Half of the participants reasoned that Google VA displays multiple options and images in a familiar UI, which helped them decide on a satisfactory answer.

For perceived naturalness, participants ranked both GazePointAR and Google VA highly with \textit{$mean_{rank}$=}1.6 (\textit{SD=}0.8) and 1.7 (\textit{SD=}0.5) respectively, followed by Google Lens (2.8; \textit{SD=}0.6). Participants generally equated naturalness to the ease with which the query was constructed (10/12). As P12 said, ``\textit{I wish I can say queries with and without pronouns, because whichever comes to mind first, that's the one I want to say.}'' Given the simplicity of the search task, P5, P11, and P12 indicated that the high specificity demanded by Google VA is not much of a concern; however, as search queries become more complex, Google VA can quickly fall behind other systems. As one example, three participants were unsure how to pronounce ``\textit{Rao's}'' so felt more comfortable saying ``\textit{this}''. While seven participants felt GazePointAR was most natural, P12 emphasized that humans are conversationally adaptable and have learned how to speak to modern VAs: ``\textit{GazePointAR was definitely the most human-like if we mean most 'natural' and 'human-like' in terms of speaking to another person; however, if we say 'natural' as in speaking to a machine, then Google Voice Assistant wins}''.

\textbf{Task completion time}. While we allowed participants to define their own stoppage mark for determining a satisfactory query answer, task time is still an interesting metric and central to information retrieval~\cite{Fox2005}. On average, the fastest completion was Google VA (\textit{avg=}26.3 secs; \textit{SD=}12.2) followed by GazePointAR (37.4 secs; \textit{SD=}11.6) then Google Lens (60.7 secs; \textit{SD=}28.3). For both Google VA and Google Lens, participants primarily spent time clicking and viewing links to find a satisfactory recipe while with GazePointAR, participants received a direct answer but were delayed by query and image processing. To form the query, Google Lens took the longest as participants had to input both an image and textual content; for both Google VA and GazePointAR, participants could form queries hands-free, which increased interaction speed.

\textbf{System usability}. Finally, for the SUS questionnaire, participants gave Google VA a higher usability score (\textit{avg=}80.0; \textit{SD=}14.3) than Google Lens (66.3; \textit{SD=}14.8) and GazePointAR (62.1; \textit{SD=}20.0)— higher is better, range is 0-100. Various factors influenced usability, including familiarity with Google suite, autonomy in choosing a satisfactory answer from Google UI, naturalness in coming up with and vocalizing queries, and task completion time.

\subsubsection{Part 2: Context-sensitive Queries} \hfill

\noindent While Part 1 explored differences between VA systems, Part 2 focuses specifically on GazePointAR and three context-sensitive queries: solving a math equation, comparing costs between items, and finding information about a celebrity (Figure~\ref{fig:part2}). We did not guide participants in how to complete the queries, so our findings are based on participants' initial instincts. For all tasks, participants chose to use gaze+speech rather than pointing as participants felt that pointing was unnecessary (7/12) and like extra work (6/12). In a few instances, participants relied on conversation history; for example, P1 asked ``\textit{How much do \textit{these} cost?}'', then, after receiving the prices of two items, they asked ``\textit{What's the cost difference?}''. Below, we report on participants' query formations and their overall reactions across tasks.



\textbf{Solving a math equation.} Interestingly, all participants constructed this query similarly: using the pronoun ``\textit{this}'', which felt most natural (9/12). As P10 said, ``\textit{the equation I wrote is right there, but I don't want to say the whole thing out loud... being able to just look and say `this' and have it read the equation is pretty useful.}'' All but one participant preferred using a context-sensitive query and pronouns compared to vocalizing the whole equation. Some participants (5/12) mentioned feeling unsure where to look to properly capture the equation during their query: ``\textit{having to keep my gaze on the equation is more difficult than a jar, since I know I have to fix my gaze, but I am not sure where I should look}'' (P3).



\textbf{Comparing costs between two items.} Unlike the math equation, participants constructed this query using two different pronouns: ten used the pronoun ``\textit{these}'' and two used ``\textit{them}''. Currently, GazePointAR only supports one pronoun per query. Five participants felt that constructing a comparative query with multiple pronouns would have felt more natural such as, ``\textit{Compare the cost of \underline{this} to \underline{that}.}'' As P1 stated, ``\textit{when there are exactly two objects, I feel like I will more likely say `\underline{this or that}' rather than `\underline{these}'}''. Similar to the math task, participants were unsure where to look to communicate intent (\textit{i.e.,} multiple object referents) with GazePointAR. Participants also reiterated wanting more system transparency to understand what GazePointAR was capturing for the context-sensitive query: ``\textit{It is impressive that it can figure out multiple objects, but it will likely be more incorrect when trying to guess multiple objects I am talking about, so I really want to know what it thought I meant}'' (P5).   


\textbf{Finding information about a celebrity.} For this task, the query construction was most varied: five used the pronoun ``\textit{this}' (\textit{e.g., ``Who is \underline{this}?''}), four ``\textit{her}'' (\textit{e.g., ``Tell me about \underline{her}.''}), and three ``\textit{she}'' (\textit{e.g., ``Who is \underline{she}?''}. Seven participants specifically mentioned how helpful pronouns were with this task: ``\textit{if you are looking at something you don't know, like a photo of a person, the only way to ask a question is by saying 'who is \underline{she}' or 'who is \underline{he}'}'' (P11). 


\subsubsection{Part 3: Design Probe and Co-design} \hfill

\noindent Finally, for Part 3, we showed five video clips of GazePointAR and then invited participants to co-brainstorm and try their own context-sensitive queries (Figure~\ref{fig:part3}). Below, we first report on reactions to our design probe and then describe participant-generated queries and how well GazePointAR performed. 
\hfill

\textbf{Reactions to design probes.} Overall, participants believed that GazePointAR has many uses, as many referents are difficult to describe in words. As P10 said: ``\textit{although I use voice assistants almost every day to play music or something, I now realize that many things I look at are difficult to clearly describe in text... since with this people can now input their environment easily, I think it will make speaking to voice assistants easier in many everyday activities.}'' P3 was surprised with the range of supported queries. Additionally, participants expressed a particular interest in the societal impact examples, such as the hazardous object clip (7/12), which shows an accessibility example where a user is asking ``\textit{Anything dangerous \underline{here}?}'' while looking ahead, and the recycling clip (4/12), which shows a user asking ``\textit{What goes in \underline{these} trash bins?}''. After viewing the hazardous object probe, P5 said, ``\textit{all you have to do is use pronouns and it can process objects in a person's field-of-view... that's great for blind people, which I really like.}'' Participants summarized that when a visual referent is either unknown or difficult to vocalize, pronouns become especially useful.

\textbf{Brainstorming and trying queries.}
For the co-design task, participants generated a total of 32 queries---see Appendix~\ref{table:queries3}---and used gaze (32/32), pointing (6/32), and conversation history (1/32). Queries in which participants used pointing gestures had pronouns ``that'' (4/6), ``there'' (1/6), and ``they'' (1/6), which were all referring to objects faraway from the user. Conversation history was used when asking follow-up questions to find more information about a celebrity. Most queries (23/32) were aimed at deriving information about an object or person, including an object's name and price, a location's distance, and a person's name and accomplishments. Other queries included foreign language translation (4/32) like ``\textit{How do you pronounce \underline{this}?}'', object comparison (3/32), and to confirm the correctness of a user's action (\textit{e.g.,} ``\textit{Can \underline{I} put \underline{this} [trash] in \underline{here} [recycling trash bin]?}'') (2/32). In analyzing pronoun usage, participants most commonly used ``\textit{this}'' (16 occurrences), followed by ``\textit{that}'' (8), ``\textit{s/he}'' or ``\textit{him/her}'' (5), and ``\textit{they}'' (1). 

GazePointAR provided a satisfactory answer for 13 of the 32 queries, including ``\textit{Who is \underline{s/he} [person]? What is \underline{his/hers} [musician] top hit?}'' and ``\textit{What's happening over \underline{there}?}''. Many of the unanswerable queries were due to lack of information, such as limitations in object recognition (\textit{e.g.,} while a object localization model can recognize a car, it does not know the make and model of the car) and missing access to information online (\textit{e.g.,} a price of an item may vary and GPT does not have access to store-specific information).

Other unanswerable queries were due to GazePointAR's inability to handle multiple pronouns in a single query (\textit{e.g.,} ``\textit{Tell me the price difference between \underline{this} and \underline{that}.}'') or past referents (\textit{e.g.,} ``\textit{Who was \underline{s/he} again?}''). Participants suggested that GazePointAR should capture gaze over time. P3 added that this will remove the need for dwelling on a referent, which will allow users to gaze more naturally and improve the system's overall usability. While P10 was in favor of this feature, they also expressed privacy concerns. P5 went even further and said GazePointAR should record objects nearby gaze to support scenarios where gaze target is not the object in question (\textit{e.g.,} ``\textit{What is the object next to \underline{that} chair}'').

\begin{figure*}[h]
  \centering
  \includegraphics[width=\linewidth]{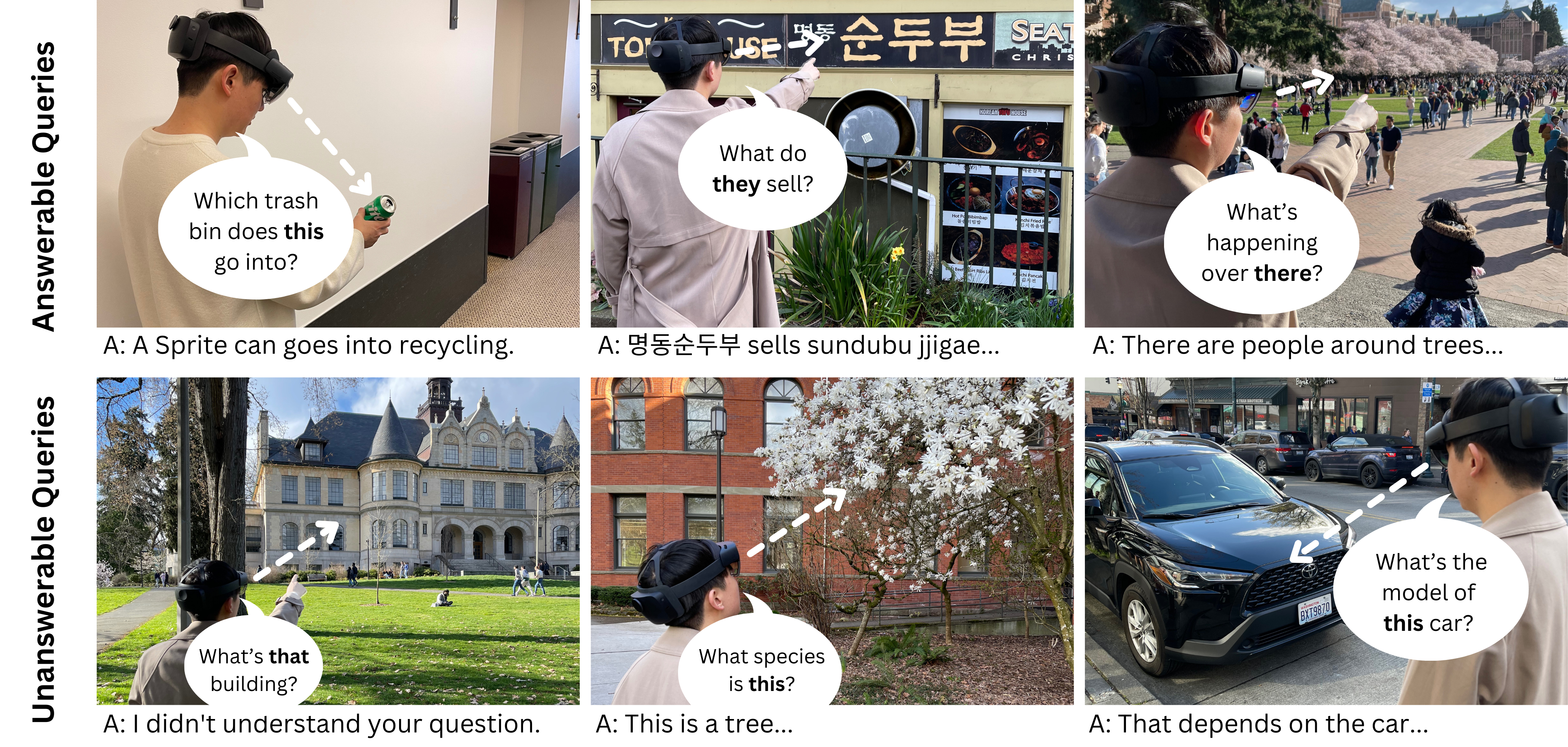}
  \caption{A subset of the scenarios participants came up with during Part 3 of Study 1. The top row shows recreations of answerable queries while the bottom rows highlights example queries that returned unsatisfactory responses.}
  \Description{A subset of queries participants came up with during Part 3 of the lab study. Answerable queries include "Which trash bin does this go into?", "What do they sell?", and "What's happening over there?". Unanswerable queries include "What's that building?", "What species is this?", and "What's the model of this car?"}
  \label{fig:additional-scenarios}
\end{figure*}

\subsection{\rev{Study 1 Summary}} \hfill

\noindent \rev{Participants appreciated GazePointAR for its simplicity, naturalness, and human-likeness. When using GazePointAR, participants mostly relied on gaze to keep the interaction hands-free and efficient, while occasionally using pointing gestures and conversation history. Participants preferred to speak pronouns, especially when referents had difficult-to-pronounce, long, or unknown names. In some cases, including pronouns in a query felt less natural (\textit{e.g.,} ``\textit{What can I make with \underline{this}?}'' vs. ``\textit{What can I make with \underline{Rao's Marinara sauce}?}''). In terms of limitations, we found that GazePointAR should support multiple pronouns, provide more answer options and explanations when answering queries, use more robust ML models, and that users could tire due to explicit gazing. Participants suggested several features for improvement: capturing gaze information over time, communicating to the user about captured images, gaze, pointing, and citations used in deriving answers, and displaying an explorable search result similar to Google.}


%% file: sections/5_system_2.tex
\section{GazePointAR Prototype 2}
\rev{Informed by Study 1 findings and our own experiences using GazePointAR, we created a second GazePointAR prototype with three advancements: first, we replaced Google Cloud Vision's Object Localization model with \textit{YOLOv8}~\cite{Jocher2023}; second, we redesigned the multimodal contextual phrase generator using prompt engineering~\cite{Reynolds2021}; third, and finally, we updated the chat completions API to leverage GPT-3.5.} We describe these advancements below and then discuss our five-day first-person diary study using GazePointAR version 2 ``\textit{in the wild}''~\cite{Rogers2017}. 

\textbf{Updating GazePointAR's object recognizer.} For the initial GazePointAR prototype, we chose Google Cloud Vision's Object Localization model, as it enabled rapid prototyping. However, a key limitation of this model is that it categorizes an object as ``\textit{packaged goods}'' if it cannot precisely identify the object, which confused both GPT-3 and our participants. In this iteration of GazePointAR, we instead employed a state-of-the-art YOLOv8 model trained on the \textit{MS COCO} dataset~\cite{Lin2015} by building a local API server using \textit{FastAPI}\footnote{https://github.com/tiangolo/fastapi} and \textit{Docker}\footnote{https://www.docker.com}, and tunneling the local API using \textit{Localtunnel}\footnote{https://github.com/localtunnel/localtunnel}. \rev{This increased the ML services' runtime to $3.75\pm0.31$ seconds (+11.28\%) and the overall runtime to $7.94\pm0.38$ seconds (+5.73\%).}

\begin{figure*}[h]
  \centering
  \includegraphics[width=\linewidth]{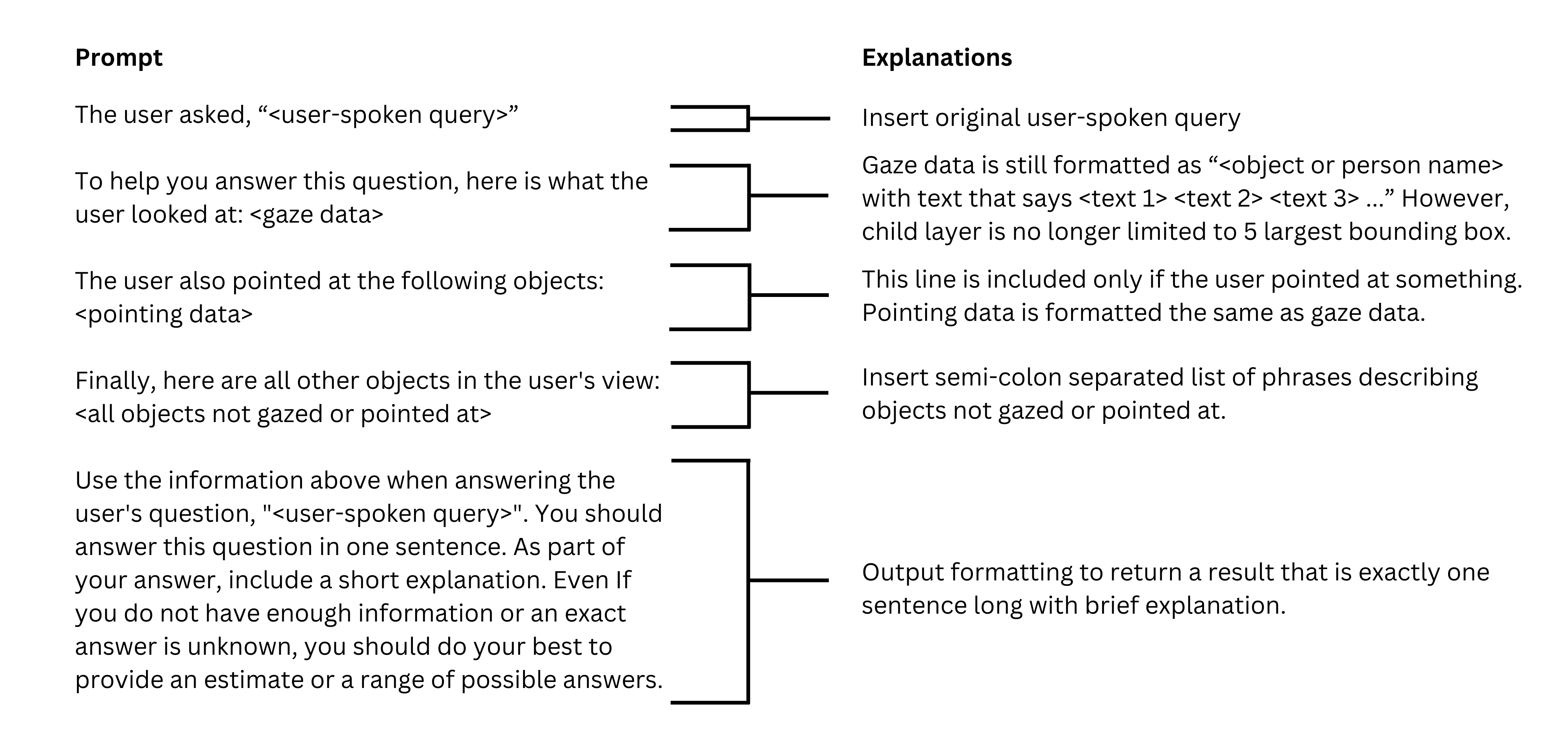}
  \caption{Engineered prompt used in GazePointAR v2}
  \Description{A figure showing the engineered prompt. It consists of the user-spoken query, gaze and pointing data, objects nearby the user's gaze, and a short paragraph asking GPT to output an answer in one sentence and include a short explanation.}
  \label{fig:prompt}
\end{figure*}

\textbf{New contextual phrase generator.} In GazePointAR v1, our phrase-generator automatically replaced query pronouns with ML results using a hierarchical heuristic model. In the revised GazePointAR prototype, we instead use prompt engineering that leverages GPT, rather than heuristics, to integrate all pieces of information together. This enabled GazePointAR v2 to support multiple pronouns, since the entirety of the original query was captured in the prompt. For example, if the user asks ``\textit{\underline{I} love \underline{this} cloth. Who designed \underline{it}?}'', rather than creating the modified query ``\textit{I love \underline{clothing with text that says [brand name]} cloth. Who designed \underline{it}?}'', GazePointAR includes the user's original query as raw information in the engineered prompt---see Figure~\ref{fig:prompt}. Note: to supply gaze and pointing gesture information, we still treat the YOLOv8 object recognition and celebrity recognition results as parent layer and OCR results as child layer to create the phrase. \rev{Additionally, as part of the prompt, we asked GazePointAR to briefly explain its answers in an attempt to enhance explainability.}

\textbf{GPT-3.5} Lastly, with the introduction of GPT-3.5, we updated GazePointAR to use \textit{gpt-3.5-turbo}, which has been trained on more up-to-date data and is more efficient than GPT-3~\cite{GPT35}.

%% file: sections/6_study_2.tex
\section{Study 2: GazePointAR Deployment}
After iterating on GazePointAR, we carried out a first-person, five-day diary study~\cite{Desjardins2021}. While informed by related first-person study methods like autoethnography~\cite{Desjardins2018, Ellis2011} and autobiography~\cite{Neustaedter2021}, we explicitly use the term ``\textit{diary study}'' as the other methods tend to span longer periods of time. The diary study enabled us to evaluate the potential of an always-available, multimodal wearable VA system in the real world. The lead researcher utilized GazePointAR v2 in their day-to-day activities while documenting their interactions. We report our process and findings.

\subsection{Procedure}
The researcher wore a Microsoft HoloLens 2 continuously running GazePointAR v2. Because GazePointAR requires an Internet connection, the HoloLens was connected to either a mobile hotspot or public Wi-Fi networks. Over five days, GazePointAR v2 was used four hours a day across various settings, including: indoor locations like homes, offices, gyms, cafes, restaurants, shopping centers, libraries, cinemas, grocery stores, and hospitals, as well as outdoor areas such as sidewalks, parks, university campuses, and public transit stations. To document their interactions, the lead researcher used HoloLens' internal video recording feature and kept a pen and notebook for journaling insights and observations.

\subsection{Findings}
In total, the lead researcher asked 48 queries, of which GazePointAR provided 20 satisfactory answers. Prompt engineering appeared to enhance the performance of GazePointAR in several ways: (1) GPT seems to recognize the importance of the user's gaze target when resolving ambiguous queries, giving it priority; (2) GPT seems to consider objects similar to the gaze target when answering queries; (3) the response is typically one sentence, and it includes a concise justification for its answer selection. Even with queries it could not answer, GazePointAR seemed to often accurately interpret user inputs and intentions, suggesting its performance was not inherently poor. For a full list of queries, see Appendix~\ref{table:diary}. Below, we present key findings including overall reflection on having an always-available context-aware VA, the types of queries asked, GazePointAR's response, and perceived limitations.

\textbf{Overall experience.} 
From simple tasks such as retrieving the rating of a new coffee shop and comparing health benefits of food items to more complicated tasks such as suggesting an allergy-friendly menu item and finding lost keys, the lead researcher set out to ``stress test'' GazePointAR v2 in the wild. They attempted to use GazePointAR naturally as an everyday assistant---looking around and posing queries as they arose. In his journal, the researcher wrote: "\textit{conversing with GazePointAR felt like a friend was tagging along, helping me}."

Perhaps the most surprising use was when, at a store, they asked: ``\textit{\underline{This} is a bit outside \underline{my} price range... can \underline{you} recommend a similar brand?}'' while looking at a piece of clothing. GazePointAR not only grasped the broader context but also identified the gaze target as clothing, determined its brand, and then recommended similar brands. However, the lead researcher recounted several instances where they felt self-conscious using GazePointAR, especially in public settings, mentioning that speaking out loud while wearing a bulky headset drew unwanted attention. This became more apparent in settings where people are typically quiet, such as libraries, hospitals, and movie theaters. \rev{Additionally, the lead researcher noted that after extended use spanning more than fifteen minutes, their eyes became tired from dwelling on referents.}

\textbf{Query Analysis.}
When analyzing the queries, we identified five categories: (1) asking for more information about a referent, such as its usage, price, and rating (21 queries); (2) asking for recommendations, such as a drink at a cafe (11); (3) asking for directions on how to proceed, such as navigating to a location or following step-by-step instructions (9); (4) asking about personal information, such as a schedule (4); and (5) asking about past actions, such as ``\textit{Did I take \underline{this} vitamin today?}'' (3). When thinking about why they used a pronoun, the lead researcher wrote ``\textit{I'm just realizing that many objects and their features are difficult to describe in words... an apple is an apple, but how do you describe how rotten it looks to a machine? Or what about a clothing stain if I want to know how to get rid of it? Also, sometimes, I don't even know the words. When I was in Chinatown, the restaurant name was only written in Chinese. How else can I ask besides saying `is \underline{this} the right place?'}'' In crafting queries, the lead researcher employed various pronouns, with ``\textit{this}'' being the most common (21 occurrences), mirroring Study 1 participants. Other pronouns include ``\textit{it}'' (6), ``\textit{that}'' (4), ``\textit{here}'' (4), ``\textit{there}'' (1), ``\textit{these}'' (1), and ``\textit{s/he}'' (1). While the lead researcher felt that the list of supported pronouns was exhaustive, 13 queries did not have pronouns in our taxonomy, and instead had first- and second person pronouns (12/13), or no pronoun at all (``\textit{What’s for sale today?}''). For multimodal input, the lead researcher found themselves relying solely on gaze rather than pointing. When asked why, they said that ``\textit{gaze was easier and hands free}''—similar reason as participants in Study 1—and that ``\textit{pointing in public spaces felt awkward.}''

Interestingly, the lead researcher often used first-person pronouns, ``\textit{I}'' (33 occurrences), ``\textit{me}'' (8) and ``\textit{my}'' (7), as well as the second-person pronoun ``\textit{you}'' (10). They observed that GazePointAR's human-like nature leads them to use full sentences in their queries, which often included first- and second-person pronouns. \rev{However, this often results in longer queries, which contradicts findings from Study 1. To justify this inconsistency, the lead researcher wrote,} ``\textit{with regular voice assistants, I feel like I'm speaking commands, while to GazePointAR, I feel like I should have conversations with it. So to Alexa, for example, I would say phrases like `price of an [item]', while to GazePointAR, I want to speak in full sentences like `Can \underline{you} tell \underline{me} the price of \underline{this} [item]?'}''. As a result, 31 queries had more than one pronoun. Finally, as part of their long queries, the lead researcher seemed to instinctively incorporate additional context. For example, when asking ``\textit{I} want to eat something light before \textit{my} commute... can \textit{you} suggest me a place?'', the lead researcher clarified their preference for a light meal and implied that the time is probably early morning. 

\begin{figure*}[h]
  \centering
  \includegraphics[width=\linewidth]{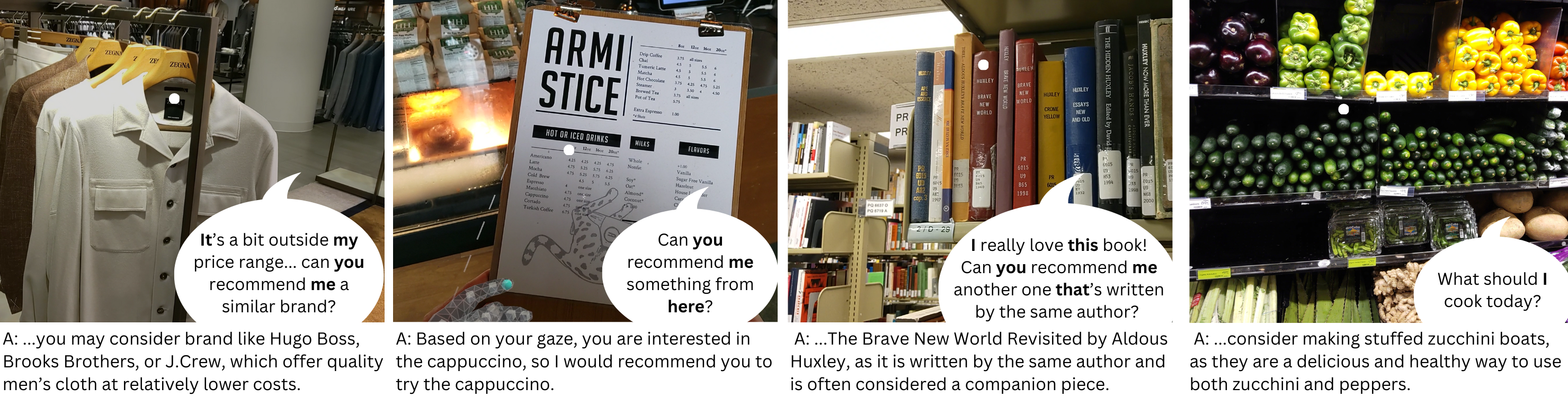}
  \caption{Example queries from the first-person diary study (Study 2) which GazePointAR answered accurately.}
  \Description{A subset of answerable queries in the first-person diary study (Study 2). On the left, a person is asking "It's a bit outside my price range... can you recommend me a similar brand?" On the middle left, a person is asking "Can you recommend me something from here?". On the middle right, a person is asking "I really love this book! Can you recommend me another one that's written by the same author?" Lastly, on the right, a person is asking "What should I cook today?”.}
  \label{fig:diaryans}
\end{figure*}

\begin{figure*}[h]
  \centering
  \includegraphics[width=\linewidth]{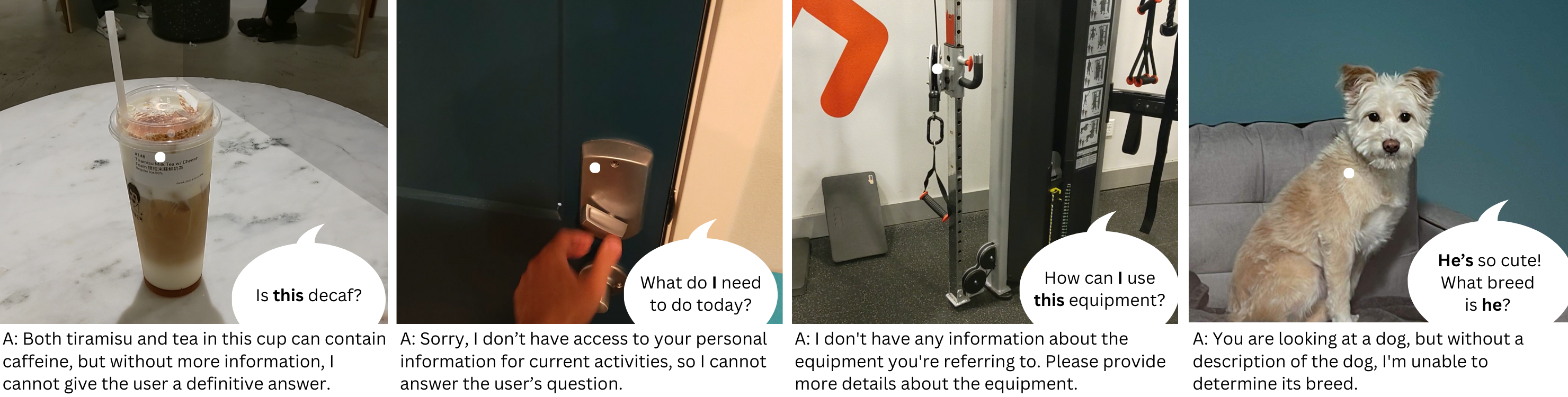}
  \caption{Example queries from the first-person diary study (Study 2) which GazePointAR answered inaccurately.}
  \Description{A subset of unanswerable queries in the first-person diary study (Study 2). On the left, a person is asking "Is this decaf?" On the middle left, a person is asking "What do I need to do today?" On the middle right, a person is asking "How can I use this equipment?" Lastly, on the right, a person is asking "He's so cute. What breed is he?”.}
  \label{fig:diaryunans}
\end{figure*}

\textbf{Query Answers.}
GazePointAR successfully addressed 20 of the 48 queries posed by the lead researcher (Figure~\ref{fig:diaryans}). For example, when asked ``\textit{Can \underline{you} recommend \underline{me} something from \underline{here}?}'', GazePointAR read text information on a menu and recommended a drink. Additionally, when asked ``\textit{I love \textit{this} cloth. Who designed \textit{it}?}'', GazePointAR not only replied with the designer's name but also provided brief information about the designer. GazePointAR even provided brief explanations, such as ``\textit{the user looked at an <object> when asking this question}''\rev{, which improved understanding of information GazePointAR captured}. In contrast, for the 28 failed queries (Figure~\ref{fig:diaryunans}), this was most commonly due to \rev{missing object category in our object recognition model} and how we capture users' gaze. For example, when asked ``\textit{How can \underline{I} use \underline{this} equipment?}'' at a gym, our object recognition model failed to recognize the different exercise equipment. Additionally, when asked ``\textit{\underline{I}’m looking for \underline{my} keys... where did \underline{I} leave \underline{it} again?}'', GazePointAR was unable to figure out the lead researcher's referent, as it does not store any information over time. \rev{Analogously, GazePointAR still had trouble with some combinations of pronouns, such as ``\textit{Which is healthier, \underline{this} or \underline{that}?}''.} To fully tackle these queries, GazePointAR needs more data, such as gaze over time and improved ML results.

\subsection{\rev{Study 2 Summary}} \hfill

\sloppy\noindent \rev{In summary, the lead researcher appreciated GazePointAR for its natural, companion-like qualities, but noted its limitations in real-world settings due to insufficient information access. GazePointAR struggled with time-dependent queries, primarily those containing referents in the past (e.g., ``\textit{\underline{That} was a really cool car! Tell \textit{me} more about \underline{it}.}''), which require gaze history or multiple referents (e.g., ``\textit{Which is the healthier option? \underline{This} or \underline{this}?}''), which require shift in gaze while speaking. Additionally, while the lead researcher employed various pronouns instinctively, he also used many first- and second-person pronouns, which led to lengthier, full-sentence queries. Furthermore, the lead researcher relied solely on gaze interaction, avoiding pointing due to the additional physical effort and its impracticality in public. Lastly, extended dwelling caused fatigue. To improve, the lead researcher suggested capturing and storing gaze data over time, and using machine learning models with more object categories.}

%% file: sections/7_discussion.tex
\section{Discussion}
\rev{By utilizing gaze, gesture, and conversation history along with an LLM, GazePointAR advances the state-of-the-art in context-aware VAs. Both the user study (Study 1) and the diary study (Study 2) highlight key benefits, including more natural query formation, always-available interaction, and human-like "assistant" qualities. Below, we discuss current challenges and future opportunities for context-aware VAs like GazePointAR.}

\textbf{\rev{Capturing gaze information over time.}} 
\rev{In both studies, some queries were unanswerable due to how GazePointAR captures gaze information---at a single moment immediately after the query has been said. Future systems should instead track gaze continuously. This would enable users to shift their gaze, promoting more natural gaze behavior and reducing fatigue from explicit gaze. Continuous gaze tracking would also let users look at multiple referents across time, and the collected gaze pattern can assist an LLM in disambiguating queries with plural pronouns (\textit{e.g.,} ``\textit{Which is cheapest among \underline{these}?}'') or multiple pronouns (\textit{e.g.,} ``\textit{Which is healthier, \underline{this} or \underline{that}?}''). Moreover, storing gaze information for later reference, even for objects no longer in sight, would be beneficial. A key challenge is to find a suitable way to present temporal gaze data in a processable format for the LLM. One solution may be to pre-process raw gaze data into features such as fixations and saccades~\cite{DavidJohn2021, Salvucci2000}, and then represent them as text for an LLM to perform referent prediction. Of course, introducing continuous gaze tracking on an AR headset may also provoke privacy concerns for both users and bystanders~\cite{Jana2013}---an additional area of future work.}

\textbf{\rev{Ensuring user autonomy in choosing an answer.}}
\rev{GazePointAR currently chooses one best answer and reads it out to the user. While this is efficient, balancing interaction speed with user autonomy in choosing answers remains a challenge. Study 1 participants preferred a Google-like UI for exploring options, while the lead researcher in Study 2 highlighted the awkwardness of having to stand still and interact with mid-air gestures in public. Moreover, the lead researcher was satisfied with GazePointAR's concise answers and explanations. A possible solution could be to first offer the top answer verbally with a brief explanation and then a Google-like UI as an option for further exploration. To further reduce cognitive load further, UI panels should be glanceable~\cite{Lu2021}, gaze-adaptive~\cite{Lindlbauer2019, Pfeuffer2021}, or show different detail levels~\cite{DiVerdi2004, Lindlbauer2019}.}

\textbf{\rev{Enhancing explainability.}}
\rev{Our study findings reinforce prior research, emphasizing the growing necessity for explainable AI (XAI) in designing everyday AI-driven experiences using wearable AR~\cite{Grubert2017, Abrash2021, Abdul2018, Xu2023}. Our initial steps included prompting an LLM to explain its responses. While this approach was quick and effective, future context-aware VAs should also visually present the captured images, user inputs, ML results, and predicted referents used to derive an answer. Again, to limit cognitive overload and UI clutter, we imagine first presenting a concise explanation followed by an option to receive more information.}

\textbf{\rev{Supporting instinctive queries.}}
\rev{Our study findings suggest that while pronouns can facilitate human-VA interaction, they are not always needed and may complicate query formation. For example, in Part 1 of Study 1, some participants preferred explicit queries such as ``\textit{What can I make with \underline{Rao's Marinara sauce}?}'' over using the pronoun ``\textit{this}''. The way individuals use pronouns in queries seems to be based on instinct and preference, which affects query ambiguity. To handle a wider range of queries, from those without pronouns to those with many, and from unambiguous to ambiguous, we integrated prompts into GazePointAR v2. This enables an LLM to process the original query, not one altered by simple heuristics, and supply ambiguous queries with relevant information. A context-aware VA should support whatever query a user thinks of first and our work shows promise in achieving this.}

\textbf{\rev{Enhancing machine learning capabilities.}} 
\rev{Other queries were unanswerable because GazePointAR's object recognition model failed to identify referents. This became more apparent in Study 2, as many real world objects are not included in YOLOv8's object categories, such as gym equipment, breeds of dogs, and types of cars. Improvements in ML algorithms~\cite{Li2023, Shen2023, supergradients2023} and the use of transformative tools like Google Lens' reverse image search or advanced multimodal LLMs such as GPT-4~\cite{GPT4} may help resolve this issue. Moreover, because many queries asked in both studies pertained to recommendations and personal data, context-aware VAs may benefit from access to personal (\textit{e.g.,} calendar) and online (\textit{e.g.,} ratings) information. Again, system designers must balance this need with the potential risks to privacy.}

\textbf{\rev{Designing a more robust study.}}
\rev{While Study 2 led to unique insights not obtainable from a lab study, it only involved the lead researcher using GazePointAR in-the-wild, which may lead to subjective results. Future research should include more participants using a context-aware VA outside the lab.}

%% file: sections/8_conclusion.tex
\section{Conclusion}
\rev{In this paper, we present GazePointAR, a context-aware multimodal VA for wearable AR capable of answering pronoun-driven ambiguous queries. In our two studies, participants appreciated GazePointAR for its naturalness and human-likeness, and ability to refer to objects that are difficult to pronounce or describe. However, participants also noted several limitations, including not collecting and storing gaze data over time, lack of autonomy and explainability, the inability to support queries with multiple or past referents, and missing object category in GazePointAR’s object recognition model. Future context-aware VAs should support innate, instinctive, and natural gaze and gesture input, as well as speech, enabling users to ask any query spontaneously.}

\begin{acks}
This work was primarily supported by an NSF Graduate Research Fellowship as well as NSF CHS \#1763199 and SCC-IRG \#2125087.
\end{acks}

%% file: sections/9_appendix.tex
\appendix
\section{Demographic Questionnaire} \label{demographic_questionnaire}
The demographic questionnaire consisted of the following questions:
\begin{enumerate}
    \item[1.] Do you have any visual impairments (e.g., blind, low vision)?
    \item[2.] Do you have any auditory impairments (e.g., deaf, hard-of-hearing)?
    \item[3.] Do you have a history of seizure?
    \item[4.] Do you have a history of epilepsy?
    \item[5.] Are you a native/bilingual/fluent English speaker?
    \item[6.] How familiar are you with voice assistant technology such as Apple Siri, Amazon Alexa, and Google Voice Assistant?
    \item[7.] List any voice assistant systems you have used before and what you used them for.
    \item[8.] How often do you use voice assistant technology?
    \item[9.] How familiar are you with augmented reality (AR) headsets and glasses, such as the Microsoft Hololens?
    \item[10.] List any AR headsets and glasses you have used before and what you used them for.
    \item[11.] How often do you use an augmented reality (AR) headset or glasses?
    \item[12.] How familiar are you with an artificial intelligence (AI) chat systems, such as chatbots and ChatGPT?
    \item[13.] List any AI chat systems you have used before and what you used them for.
    \item[14.] How often do you use an AI chat system?
\end{enumerate}

\section{Queries}

\begin{table*}[h]
  \centering
  \begin{tabular}{@{\extracolsep{\fill}}lll}
    \toprule
    
    P\# & System & Query \\
    
    \midrule
    
    \multirow{3}{*}{P1} 
    & Google Voice Assistant & What can I make with Rao's Marinara sauce? \\
    & Google Lens & Image + ``Recipe'' \\
    & GazePointAR & What can I make with this? \\
    \hline

    \multirow{3}{*}{P2}
    & Google Voice Assistant & Find me a recipe that uses Rao's homemade Marinara sauce. \\
    & Google Lens & Image + ``Recipe'' \\
    & GazePointAR & Find me a good recipe to make with this. \\
    \hline

    \multirow{3}{*}{P3}
    & Google Voice Assistant & Recipes using Rao's homemade Marinara sauce 24 ounces. \\
    & Google Lens & Image + ``Recipe using'' \\
    & GazePointAR & Find me recipes using this. \\
    \hline

    \multirow{3}{*}{P4}
    & Google Voice Assistant & Find me a recipe using Rao's Marinara sauce. \\
    & Google Lens & Image + ``Recipe'' \\
    & GazePointAR & Find me a recipe using this. \\
    \hline

    \multirow{3}{*}{P5}
    & Google Voice Assistant & Recipe with Rao's homemade Marinara sauce. \\
    & Google Lens & Image + ``Recipe using'' \\
    & GazePointAR & Find me a recipe with this. \\
    \hline

    \multirow{3}{*}{P6}
    & Google Voice Assistant & Recipe using Rao's homemade Marinara sauce. \\
    & Google Lens & Image + ``Recipe'' \\
    & GazePointAR & Find a recipe using this. \\
    \hline

    \multirow{3}{*}{P7}
    & Google Voice Assistant & Find me a recipe including Rao's homemade Marinara sauce. \\
    & Google Lens & Image + ``Recipe'' \\
    & GazePointAR & Find me a recipe using this ingredient. \\
    \hline

    \multirow{3}{*}{P8}
    & Google Voice Assistant & Find me a recipe with Rao's homemade Marinara. \\
    & Google Lens & Image + ``Recipe'' \\
    & GazePointAR & Tell me a recipe that use this. \\
    \hline

    \multirow{3}{*}{P9}
    & Google Voice Assistant & Can you search for a recipe that is using Rao's homemade Marinara? \\
    & Google Lens & Image + ``Recipe'' \\
    & GazePointAR & Can you give me the recipe that is using this? \\
    \hline

    \multirow{3}{*}{P10}
    & Google Voice Assistant & Search for a recipe using Rao's homemade Marinara sauce. \\
    & Google Lens & Image + ``Recipe'' \\
    & GazePointAR & Search for a recipe using this. \\
    \hline

    \multirow{3}{*}{P11}
    & Google Voice Assistant & Can you find me a recipe that is using Rao's homemade Marinara? \\
    & Google Lens & Image + ``Recipe using'' \\
    & GazePointAR & Find me the recipe with this. \\
    \hline

    \multirow{3}{*}{P12}
    & Google Voice Assistant & Recipe using Rao's Marinara sauce. \\
    & Google Lens & Image + ``Recipe'' \\
    & GazePointAR & Recipe using this. \\
    
    \bottomrule
  \end{tabular}
  \caption{User-spoken Queries in Part 1 of the Study}
  \label{table:queries1}
\end{table*}

\begin{table*}[h]
  \centering
  \begin{tabular}{@{\extracolsep{\fill}}lll}
    \toprule
    
    Task & P\# & Query \\
    
    \midrule
    
    \multirow{12}{*}{Math Task}
    & P1 & Is this equation correct? \\
    & P2 & Did I do this equation right? \\
    & P3 & Is this correct? \\
    & P4 & Is this equation correct? \\
    & P5 & What's the answer to this equation? \\
    & P6 & Is this equation correct? \\
    & P7 & Is this correct? \\
    & P8 & Is this correct? \\
    & P9 & Is this equation correct? \\
    & P10 & Is this correct? \\
    & P11 & Tell me if this is correct. \\
    & P12 & Is this correct? \\

    \hline

    \multirow{12}{*}{Price Difference Task}
    & P1 & How much do these cost? \\
    & P2 & Which of these is more expensive, and by how much? \\
    & P3 & Can you compare the price between these two? \\
    & P4 & What's the price difference between these two items? \\
    & P5 & Find me the difference in costs between these two items. \\
    & P6 & What's the price difference between these? \\
    & P7 & What is the difference in price between these? \\
    & P8 & How much is the price difference between these? \\
    & P9 & What's the price difference between them? \\
    & P10 & What's the price difference between these? \\
    & P11 & Tell me the price difference between them. \\
    & P12 & What's the price difference between these? \\

    \hline

    \multirow{12}{*}{Celebrity Task}
    & P1 & Who is she? \\
    & P2 & Can you tell me more information about her? \\
    & P3 & Who is this person? \\
    & P4 & Who is this person? \\
    & P5 & Find me more information about this person. \\
    & P6 & Tell me about her. \\
    & P7 & Who is this? \\
    & P8 & Find me more information about her. \\
    & P9 & Who is she? \\
    & P10 & Who is she? \\
    & P11 & Tell me more about her. \\
    & P12 & Who is this? \\
    
    \bottomrule
  \end{tabular}
  \caption{User-spoken Queries in Part 2 of the Study}
  \label{table:queries2}
\end{table*}

\begin{table*}[h]
  \centering
  \begin{tabular}{@{\extracolsep{\fill}}lll}
    \toprule
    
    P\# & Query & Satisfactory? \\
    
    \midrule
    
    \multirow{1}{*}{P1} 
    & Tell \textit{me} the price difference between \emph{this} and \emph{that}. & No \\

    \hline

    \multirow{2}{*}{P2} 
    & Did \textit{I} solve \emph{this} [complex calculus problem] correctly? & No \\
    & How far away am \textit{I} from \textit{my} house? & No \\

    \hline

    \multirow{3}{*}{P3} 
    & Which trash bin does \emph{this} [trash] go into? & Yes \\
    & Can \textit{I} put \emph{this} [trash] in any of \emph{these} [trash bins]? & Yes \\
    & Can \textit{I} put \textit{this} [trash] in \textit{here} [recyling trash bin]? & No \\

    \hline

    \multirow{3}{*}{P4} 
    & Can \textit{you} explain \emph{that} [diagram in a classroom]? & No \\
    & What can \textit{I} use to clean \emph{this} [stain on a surface]? & No \\
    & Can \textit{you} translate what \emph{she} [foreigner] is saying? & No \\

    \hline

    \multirow{4}{*}{P5} 
    & A child constantly asking ``What's \emph{this}?'' & Yes \\
    & A blind person asking ``What did \emph{s/he} [speaker] point to?'' & No \\
    & Tell \textit{me} the price difference between \emph{this} and \emph{this}. & No \\
    & What is in the box sitting on top of \textit{that} chair? & No \\

    \hline

    \multirow{7}{*}{P6} 
    & Tell \textit{me} more about \emph{that} building. & No \\
    & Who made \emph{that} [car]? & No \\
    & What species is \textit{this} [plant]? & No \\
    & A blind person can ask ``Who is \emph{that} on TV?'' & Yes \\
    & Who wrote \emph{it} [book]? Tell \textit{me} more about the author. & Yes \\
    & Who are \emph{those} people? & No \\
    & What's happening over \emph{there}? & Yes \\
    & What is the object next to \textit{that} [an object I know]? & No \\

    \hline

    \multirow{2}{*}{P7} 
    & How do \textit{you} pronounce \emph{this} [foreign or complex word]? & Yes \\
    & How do \textit{you} translate \emph{this} [foreign or complex word]? & Yes \\

    \hline

    \multirow{2}{*}{P9} 
    & Who is \emph{s/he} [person]? & Yes \\
    & What is \emph{his/hers} [musician] top hit? & Yes \\

    \hline

    \multirow{1}{*}{P10} 
    & Who is left of \textit{him/her}? & No \\
    & What do \textit{they} sell? & Yes \\

    \hline

    \multirow{1}{*}{P11} 
    & What does \emph{this} mean [foreign language]? & Yes \\

    \hline

    \multirow{4}{*}{P12} 
    & Can \textit{you} tell me more about \emph{this} [unknown objects]? & Yes \\
    & Who was \emph{s/he} [celebrity] again? & No \\
    & Compare \emph{this} to \emph{that} thing from before [an object I saw a few seconds ago]. & No \\
    & When did \textit{I} do \textit{this} [activity]? & No \\

    \hline
    \end{tabular}
    \caption{User-spoken Queries in Part 3 of the Study}
  \label{table:queries3}
\end{table*}

\begin{table*}[h]
  \centering
  \begin{tabular}{@{\extracolsep{\fill}}lll}
    \toprule
    
    Location & Query & Satisfactory? \\
    
    \midrule
    
    \multirow{12}{*}{Home} 
    & What do \textit{I} need to do today? & No\\
    & \textit{My} plant seems to be dying. What can \textit{I} do for \textit{it}? & Yes \\
    & \textit{I'm} done with \textit{this}. Where can \textit{I} buy another one? & Yes \\
    & What should \textit{I} eat today? & Yes \\
    & \textit{I} want to grab coffee on the way... suggest \textit{me} a cafe nearby. & No \\
    & What's the best settings on \textit{this} [coffee] machine? & No \\
    & Does \textit{this} look spoiled? & No \\
    & \textit{I}'m looking for \textit{my} keys... where did \textit{I} leave \textit{it} again? & No \\
    & Can \textit{you} let \textit{me} know when \textit{you} find \textit{my} keys? & No \\
    & \textit{I} stained \textit{this} clothing... how can \textit{I} remove \textit{it}? & No \\
    & Did \textit{I} take \textit{this} vitamin today? & No \\
    & Did \textit{I} turn off the stove? & No \\

    \hline

    \multirow{2}{*}{Work} 
    & What's \textit{my} agenda for today? & No \\
    & \textit{I} want to eat something light before \textit{my} commute... can \textit{you} suggest \textit{me} a place? & No \\

    \hline

    \multirow{3}{*}{Gym} 
    & How can \textit{I} use \textit{this} equipment? & No \\
    & \textit{I}'m working on legs today... can \textit{you} recommend a workout plan? & Yes \\
    & What should \textit{I} eat post workout? & Yes \\

    \hline

    \multirow{4}{*}{Cafe} 
    & How well rated is \textit{this} coffee shop? & Yes \\
    & Can \textit{you} recommend \textit{me} something from \textit{here}? & Yes \\
    & After \textit{I}'m done, where should \textit{I} toss \textit{this}? & No \\
    & Is \textit{this} decaf? & No \\

    \hline

    \multirow{3}{*}{Restaurant} 
    & Can \textit{you} recommend something from \textit{here}? & Yes \\
    & \textit{I}'m allergic to \textit{that}... can \textit{you} recommend something else? & Yes \\
    & \textit{I} love \textit{this} dish! How can \textit{I} make \textit{this} from home? & No \\

    \hline

    \multirow{4}{*}{Shopping Mall} 
    & What's \textit{this} store known for? & Yes \\
    & Which of \textit{these} stores should \textit{I} visit? & Yes \\
    & \textit{I} love \textit{this} cloth. Who designed \textit{it}? & Yes \\
    & \textit{It}'s a bit outside \textit{my} price range... can \textit{you} recommend \textit{me} a similar brand? & Yes \\

    \hline

    \multirow{2}{*}{Library} 
    & \textit{I} really love \textit{this} book. Can \textit{you} recommend another book by the same author? & Yes \\
    & What's one latest book \textit{you} can recommend that \textit{I} read? & Yes \\

    \hline

    \multirow{1}{*}{Movie Theater} 
    & Is \textit{this} a good movie? & Yes \\
    & Tell \textit{me} the history behind \textit{this} scene & No \\

    \hline

    \multirow{1}{*}{Grocery Store} 
    & What should \textit{I} cook today? & Yes \\
    & What's for sale today? & No \\
    & Which is the healthier option? \textit{This} or \textit{this}? & Yes \\
    & Anything \textit{I}'m missing \textit{here}? & No \\

    \hline

    \multirow{2}{*}{Hospital} 
    & Pull up \textit{my} appointment details. & No \\
    & Do \textit{I} have to be anywhere after \textit{this}? & No \\

    \hline

    \multirow{3}{*}{Park} 
    & What [dog] breed is \textit{s/he}? & No \\
    & Can \textit{you} tell \textit{me} more about \textit{that} plant? & No \\
    & Can \textit{I} buy \textit{this} plant from somewhere? & No \\

    \hline

    \multirow{2}{*}{University Campus} 
    & \textit{I}'m supposed to meet a friend from [location]. How do \textit{I} get \textit{there}? & No \\
    & When was \textit{this} building built? & Yes \\

    \hline

    \multirow{2}{*}{Public Transit Station} 
    & \textit{I}'m trying to get to [location]. Is \textit{this} the bus \textit{I} should take? & No \\
    & Where should \textit{I} go from \textit{here}? & No \\

    \hline

    \multirow{3}{*}{Sidewalk} 
    & \textit{That} was a really cool car! Tell \textit{me} more about \textit{it}. & No \\
    & When does \textit{that} store close? & Yes \\
    & Is \textit{this} the right place [store I am trying to reach]? & No \\

    \hline
    \end{tabular}
    \caption{User-spoken Queries in the First-Person Diary Study}
    \label{table:diary}
\end{table*}

%% file: paper.bbl

\begin{thebibliography}{92}


\ifx \showCODEN    \undefined \def \showCODEN     #1{\unskip}     \fi
\ifx \showDOI      \undefined \def \showDOI       #1{#1}\fi
\ifx \showISBNx    \undefined \def \showISBNx     #1{\unskip}     \fi
\ifx \showISBNxiii \undefined \def \showISBNxiii  #1{\unskip}     \fi
\ifx \showISSN     \undefined \def \showISSN      #1{\unskip}     \fi
\ifx \showLCCN     \undefined \def \showLCCN      #1{\unskip}     \fi
\ifx \shownote     \undefined \def \shownote      #1{#1}          \fi
\ifx \showarticletitle \undefined \def \showarticletitle #1{#1}   \fi
\ifx \showURL      \undefined \def \showURL       {\relax}        \fi
\providecommand\bibfield[2]{#2}
\providecommand\bibinfo[2]{#2}
\providecommand\natexlab[1]{#1}
\providecommand\showeprint[2][]{arXiv:#2}

\bibitem[Abdolrahmani et~al\mbox{.}(2021)]%
        {Abdolrahmani2021}
\bibfield{author}{\bibinfo{person}{Ali Abdolrahmani}, \bibinfo{person}{Maya
  Howes~Gupta}, \bibinfo{person}{Mei-Lian Vader}, \bibinfo{person}{Ravi Kuber},
  {and} \bibinfo{person}{Stacy Branham}.} \bibinfo{year}{2021}\natexlab{}.
\newblock \showarticletitle{Towards More Transactional Voice Assistants:
  Investigating the Potential for a Multimodal Voice-Activated Indoor
  Navigation Assistant for Blind and Sighted Travelers}. In
  \bibinfo{booktitle}{\emph{Proceedings of the 2021 CHI Conference on Human
  Factors in Computing Systems}} (Yokohama, Japan) \emph{(\bibinfo{series}{CHI
  '21})}. \bibinfo{publisher}{Association for Computing Machinery},
  \bibinfo{address}{New York, NY, USA}, Article \bibinfo{articleno}{495},
  \bibinfo{numpages}{16}~pages.
\newblock
\showISBNx{9781450380966}
\urldef\tempurl%
\url{https://doi.org/10.1145/3411764.3445638}
\showDOI{\tempurl}


\bibitem[Abdul et~al\mbox{.}(2018)]%
        {Abdul2018}
\bibfield{author}{\bibinfo{person}{Ashraf Abdul}, \bibinfo{person}{Jo
  Vermeulen}, \bibinfo{person}{Danding Wang}, \bibinfo{person}{Brian~Y. Lim},
  {and} \bibinfo{person}{Mohan Kankanhalli}.} \bibinfo{year}{2018}\natexlab{}.
\newblock \showarticletitle{Trends and Trajectories for Explainable,
  Accountable and Intelligible Systems: An HCI Research Agenda}. In
  \bibinfo{booktitle}{\emph{Proceedings of the 2018 CHI Conference on Human
  Factors in Computing Systems}} (Montreal QC, Canada)
  \emph{(\bibinfo{series}{CHI '18})}. \bibinfo{publisher}{Association for
  Computing Machinery}, \bibinfo{address}{New York, NY, USA},
  \bibinfo{pages}{1–18}.
\newblock
\showISBNx{9781450356206}
\urldef\tempurl%
\url{https://doi.org/10.1145/3173574.3174156}
\showDOI{\tempurl}


\bibitem[Abrash(2021)]%
        {Abrash2021}
\bibfield{author}{\bibinfo{person}{Michael Abrash}.}
  \bibinfo{year}{2021}\natexlab{}.
\newblock \showarticletitle{Creating the Future: Augmented Reality, the next
  Human-Machine Interface}. In \bibinfo{booktitle}{\emph{2021 IEEE
  International Electron Devices Meeting (IEDM)}}.
  \bibinfo{pages}{1.2.1--1.2.11}.
\newblock
\urldef\tempurl%
\url{https://doi.org/10.1109/IEDM19574.2021.9720526}
\showDOI{\tempurl}


\bibitem[Aftab(2019)]%
        {Aftab2019}
\bibfield{author}{\bibinfo{person}{Abdul~Rafey Aftab}.}
  \bibinfo{year}{2019}\natexlab{}.
\newblock \showarticletitle{Multimodal Driver Interaction with Gesture, Gaze
  and Speech}. In \bibinfo{booktitle}{\emph{2019 International Conference on
  Multimodal Interaction}} (Suzhou, China) \emph{(\bibinfo{series}{ICMI '19})}.
  \bibinfo{publisher}{Association for Computing Machinery},
  \bibinfo{address}{New York, NY, USA}, \bibinfo{pages}{487–492}.
\newblock
\showISBNx{9781450368605}
\urldef\tempurl%
\url{https://doi.org/10.1145/3340555.3356093}
\showDOI{\tempurl}


\bibitem[Aharon et~al\mbox{.}(2023)]%
        {supergradients2023}
\bibfield{author}{\bibinfo{person}{Shay Aharon},
  \bibinfo{person}{{Louis-Dupont}}, \bibinfo{person}{{Ofri Masad}},
  \bibinfo{person}{Kate Yurkova}, \bibinfo{person}{{Lotem Fridman}},
  \bibinfo{person}{{Lkdci}}, \bibinfo{person}{Eugene Khvedchenya},
  \bibinfo{person}{Ran Rubin}, \bibinfo{person}{Natan Bagrov},
  \bibinfo{person}{Borys Tymchenko}, \bibinfo{person}{Tomer Keren},
  \bibinfo{person}{Alexander Zhilko}, {and} \bibinfo{person}{{Eran-Deci}}.}
  \bibinfo{year}{2023}\natexlab{}.
\newblock \bibinfo{title}{Super-Gradients}.
\newblock
\newblock
\urldef\tempurl%
\url{https://doi.org/10.5281/ZENODO.7789328}
\showDOI{\tempurl}


\bibitem[Ammari et~al\mbox{.}(2019)]%
        {Ammari2019}
\bibfield{author}{\bibinfo{person}{Tawfiq Ammari}, \bibinfo{person}{Jofish
  Kaye}, \bibinfo{person}{Janice~Y. Tsai}, {and} \bibinfo{person}{Frank
  Bentley}.} \bibinfo{year}{2019}\natexlab{}.
\newblock \showarticletitle{Music, Search, and IoT: How People (Really) Use
  Voice Assistants}.
\newblock \bibinfo{journal}{\emph{ACM Trans. Comput.-Hum. Interact.}}
  \bibinfo{volume}{26}, \bibinfo{number}{3}, Article \bibinfo{articleno}{17}
  (\bibinfo{date}{apr} \bibinfo{year}{2019}), \bibinfo{numpages}{28}~pages.
\newblock
\showISSN{1073-0516}
\urldef\tempurl%
\url{https://doi.org/10.1145/3311956}
\showDOI{\tempurl}


\bibitem[AWS(2023)]%
        {AmazonRekognition}
\bibfield{author}{\bibinfo{person}{Amazon AWS}.}
  \bibinfo{year}{2023}\natexlab{}.
\newblock \bibinfo{title}{Amazon Rekognition}.
\newblock
\newblock
\urldef\tempurl%
\url{https://aws.amazon.com/rekognition/}
\showURL{%
\tempurl}


\bibitem[Bentley et~al\mbox{.}(2018)]%
        {Bentley2018}
\bibfield{author}{\bibinfo{person}{Frank Bentley}, \bibinfo{person}{Chris
  Luvogt}, \bibinfo{person}{Max Silverman}, \bibinfo{person}{Rushani
  Wirasinghe}, \bibinfo{person}{Brooke White}, {and} \bibinfo{person}{Danielle
  Lottridge}.} \bibinfo{year}{2018}\natexlab{}.
\newblock \showarticletitle{Understanding the Long-Term Use of Smart Speaker
  Assistants}.
\newblock \bibinfo{journal}{\emph{Proc. ACM Interact. Mob. Wearable Ubiquitous
  Technol.}} \bibinfo{volume}{2}, \bibinfo{number}{3}, Article
  \bibinfo{articleno}{91} (\bibinfo{date}{sep} \bibinfo{year}{2018}),
  \bibinfo{numpages}{24}~pages.
\newblock
\urldef\tempurl%
\url{https://doi.org/10.1145/3264901}
\showDOI{\tempurl}


\bibitem[Bolt(1980)]%
        {Bolt1980}
\bibfield{author}{\bibinfo{person}{Richard~A. Bolt}.}
  \bibinfo{year}{1980}\natexlab{}.
\newblock \showarticletitle{“Put-That-There”: Voice and Gesture at the
  Graphics Interface}. In \bibinfo{booktitle}{\emph{Proceedings of the 7th
  Annual Conference on Computer Graphics and Interactive Techniques}} (Seattle,
  Washington, USA) \emph{(\bibinfo{series}{SIGGRAPH '80})}.
  \bibinfo{publisher}{Association for Computing Machinery},
  \bibinfo{address}{New York, NY, USA}, \bibinfo{pages}{262–270}.
\newblock
\showISBNx{0897910214}
\urldef\tempurl%
\url{https://doi.org/10.1145/800250.807503}
\showDOI{\tempurl}


\bibitem[Braun and Clarke(2006)]%
        {Braun2006}
\bibfield{author}{\bibinfo{person}{Virginia Braun} {and}
  \bibinfo{person}{Victoria Clarke}.} \bibinfo{year}{2006}\natexlab{}.
\newblock \showarticletitle{Using thematic analysis in psychology}.
\newblock \bibinfo{journal}{\emph{Qualitative Research in Psychology}}
  \bibinfo{volume}{3}, \bibinfo{number}{2} (\bibinfo{year}{2006}),
  \bibinfo{pages}{77--101}.
\newblock
\urldef\tempurl%
\url{https://doi.org/10.1191/1478088706qp063oa}
\showDOI{\tempurl}
\showeprint{https://www.tandfonline.com/doi/pdf/10.1191/1478088706qp063oa}


\bibitem[Braun and Clarke(2019)]%
        {Braun2019}
\bibfield{author}{\bibinfo{person}{Virginia Braun} {and}
  \bibinfo{person}{Victoria Clarke}.} \bibinfo{year}{2019}\natexlab{}.
\newblock \showarticletitle{Reflecting on reflexive thematic analysis}.
\newblock \bibinfo{journal}{\emph{Qualitative Research in Sport, Exercise and
  Health}} \bibinfo{volume}{11}, \bibinfo{number}{4} (\bibinfo{year}{2019}),
  \bibinfo{pages}{589--597}.
\newblock
\urldef\tempurl%
\url{https://doi.org/10.1080/2159676X.2019.1628806}
\showDOI{\tempurl}
\showeprint{https://doi.org/10.1080/2159676X.2019.1628806}


\bibitem[Brooke(1996)]%
        {Brooke1996}
\bibfield{author}{\bibinfo{person}{J.~B. Brooke}.}
  \bibinfo{year}{1996}\natexlab{}.
\newblock \showarticletitle{SUS: A 'Quick and Dirty' Usability Scale}.
\newblock


\bibitem[Byron and Allen(1998)]%
        {Byron1998}
\bibfield{author}{\bibinfo{person}{Donna~K. Byron} {and}
  \bibinfo{person}{James~F. Allen}.} \bibinfo{year}{1998}\natexlab{}.
\newblock \showarticletitle{Resolving Demonstrative Anaphora in the TRAINS93
  Corpus}.
\newblock


\bibitem[Cao and Daumé(2021)]%
        {Cao2021}
\bibfield{author}{\bibinfo{person}{Yang~Trista Cao} {and} \bibinfo{person}{III
  Daumé, Hal}.} \bibinfo{year}{2021}\natexlab{}.
\newblock \showarticletitle{{Toward Gender-Inclusive Coreference Resolution: An
  Analysis of Gender and Bias Throughout the Machine Learning Lifecycle*}}.
\newblock \bibinfo{journal}{\emph{Computational Linguistics}}
  \bibinfo{volume}{47}, \bibinfo{number}{3} (\bibinfo{date}{11}
  \bibinfo{year}{2021}), \bibinfo{pages}{615--661}.
\newblock
\showISSN{0891-2017}
\urldef\tempurl%
\url{https://doi.org/10.1162/coli_a_00413}
\showDOI{\tempurl}
\showeprint{https://direct.mit.edu/coli/article-pdf/47/3/615/1971880/coli\_a\_00413.pdf}


\bibitem[Chatterjee et~al\mbox{.}(2015)]%
        {Chatterjee2015}
\bibfield{author}{\bibinfo{person}{Ishan Chatterjee}, \bibinfo{person}{Robert
  Xiao}, {and} \bibinfo{person}{Chris Harrison}.}
  \bibinfo{year}{2015}\natexlab{}.
\newblock \showarticletitle{Gaze+Gesture: Expressive, Precise and Targeted
  Free-Space Interactions}. In \bibinfo{booktitle}{\emph{Proceedings of the
  2015 ACM on International Conference on Multimodal Interaction}} (Seattle,
  Washington, USA) \emph{(\bibinfo{series}{ICMI '15})}.
  \bibinfo{publisher}{Association for Computing Machinery},
  \bibinfo{address}{New York, NY, USA}, \bibinfo{pages}{131–138}.
\newblock
\showISBNx{9781450339124}
\urldef\tempurl%
\url{https://doi.org/10.1145/2818346.2820752}
\showDOI{\tempurl}


\bibitem[Cloud(2023)]%
        {GoogleCloudVision}
\bibfield{author}{\bibinfo{person}{Google Cloud}.}
  \bibinfo{year}{2023}\natexlab{}.
\newblock \bibinfo{title}{Vision AI}.
\newblock
\newblock
\urldef\tempurl%
\url{https://cloud.google.com/vision}
\showURL{%
\tempurl}


\bibitem[Cohen et~al\mbox{.}(1997)]%
        {Cohen1997}
\bibfield{author}{\bibinfo{person}{Philip~R. Cohen}, \bibinfo{person}{Michael
  Johnston}, \bibinfo{person}{David McGee}, \bibinfo{person}{Sharon Oviatt},
  \bibinfo{person}{Jay Pittman}, \bibinfo{person}{Ira Smith},
  \bibinfo{person}{Liang Chen}, {and} \bibinfo{person}{Josh Clow}.}
  \bibinfo{year}{1997}\natexlab{}.
\newblock \showarticletitle{QuickSet: Multimodal Interaction for Distributed
  Applications}. In \bibinfo{booktitle}{\emph{Proceedings of the Fifth ACM
  International Conference on Multimedia}} (Seattle, Washington, USA)
  \emph{(\bibinfo{series}{MULTIMEDIA '97})}. \bibinfo{publisher}{Association
  for Computing Machinery}, \bibinfo{address}{New York, NY, USA},
  \bibinfo{pages}{31–40}.
\newblock
\showISBNx{0897919912}
\urldef\tempurl%
\url{https://doi.org/10.1145/266180.266328}
\showDOI{\tempurl}


\bibitem[Corbett and Chang(1983)]%
        {Corbett1983}
\bibfield{author}{\bibinfo{person}{Albert~T Corbett} {and}
  \bibinfo{person}{Frederick~R Chang}.} \bibinfo{year}{1983}\natexlab{}.
\newblock \showarticletitle{Pronoun disambiguation: Accessing potential
  antecedents}.
\newblock \bibinfo{journal}{\emph{Memory \& Cognition}} \bibinfo{volume}{11},
  \bibinfo{number}{3} (\bibinfo{year}{1983}), \bibinfo{pages}{283--294}.
\newblock
\urldef\tempurl%
\url{https://doi.org/10.3758/BF03196975}
\showURL{%
\tempurl}


\bibitem[Dahlb\"{a}ck et~al\mbox{.}(1993)]%
        {Dahlback1993}
\bibfield{author}{\bibinfo{person}{Nils Dahlb\"{a}ck}, \bibinfo{person}{Arne
  J\"{o}nsson}, {and} \bibinfo{person}{Lars Ahrenberg}.}
  \bibinfo{year}{1993}\natexlab{}.
\newblock \showarticletitle{Wizard of Oz Studies: Why and How}. In
  \bibinfo{booktitle}{\emph{Proceedings of the 1st International Conference on
  Intelligent User Interfaces}} (Orlando, Florida, USA)
  \emph{(\bibinfo{series}{IUI '93})}. \bibinfo{publisher}{Association for
  Computing Machinery}, \bibinfo{address}{New York, NY, USA},
  \bibinfo{pages}{193–200}.
\newblock
\showISBNx{0897915569}
\urldef\tempurl%
\url{https://doi.org/10.1145/169891.169968}
\showDOI{\tempurl}


\bibitem[David-John et~al\mbox{.}(2021)]%
        {DavidJohn2021}
\bibfield{author}{\bibinfo{person}{Brendan David-John},
  \bibinfo{person}{Candace Peacock}, \bibinfo{person}{Ting Zhang},
  \bibinfo{person}{T.~Scott Murdison}, \bibinfo{person}{Hrvoje Benko}, {and}
  \bibinfo{person}{Tanya~R. Jonker}.} \bibinfo{year}{2021}\natexlab{}.
\newblock \showarticletitle{Towards Gaze-Based Prediction of the Intent to
  Interact in Virtual Reality}. In \bibinfo{booktitle}{\emph{ACM Symposium on
  Eye Tracking Research and Applications}} (Virtual Event, Germany)
  \emph{(\bibinfo{series}{ETRA '21 Short Papers})}.
  \bibinfo{publisher}{Association for Computing Machinery},
  \bibinfo{address}{New York, NY, USA}, Article \bibinfo{articleno}{2},
  \bibinfo{numpages}{7}~pages.
\newblock
\showISBNx{9781450383455}
\urldef\tempurl%
\url{https://doi.org/10.1145/3448018.3458008}
\showDOI{\tempurl}


\bibitem[Desjardins and Ball(2018)]%
        {Desjardins2018}
\bibfield{author}{\bibinfo{person}{Audrey Desjardins} {and}
  \bibinfo{person}{Aubree Ball}.} \bibinfo{year}{2018}\natexlab{}.
\newblock \showarticletitle{Revealing Tensions in Autobiographical Design in
  HCI}. In \bibinfo{booktitle}{\emph{Proceedings of the 2018 Designing
  Interactive Systems Conference}} (Hong Kong, China)
  \emph{(\bibinfo{series}{DIS '18})}. \bibinfo{publisher}{Association for
  Computing Machinery}, \bibinfo{address}{New York, NY, USA},
  \bibinfo{pages}{753–764}.
\newblock
\showISBNx{9781450351980}
\urldef\tempurl%
\url{https://doi.org/10.1145/3196709.3196781}
\showDOI{\tempurl}


\bibitem[Desjardins et~al\mbox{.}(2021)]%
        {Desjardins2021}
\bibfield{author}{\bibinfo{person}{Audrey Desjardins}, \bibinfo{person}{Oscar
  Tomico}, \bibinfo{person}{Andr\'{e}s Lucero}, \bibinfo{person}{Marta~E.
  Cecchinato}, {and} \bibinfo{person}{Carman Neustaedter}.}
  \bibinfo{year}{2021}\natexlab{}.
\newblock \showarticletitle{Introduction to the Special Issue on First-Person
  Methods in HCI}.
\newblock \bibinfo{journal}{\emph{ACM Trans. Comput.-Hum. Interact.}}
  \bibinfo{volume}{28}, \bibinfo{number}{6}, Article \bibinfo{articleno}{37}
  (\bibinfo{date}{dec} \bibinfo{year}{2021}), \bibinfo{numpages}{12}~pages.
\newblock
\showISSN{1073-0516}
\urldef\tempurl%
\url{https://doi.org/10.1145/3492342}
\showDOI{\tempurl}


\bibitem[Diessel and Coventry(2020)]%
        {Diessel2020}
\bibfield{author}{\bibinfo{person}{Holger Diessel} {and}
  \bibinfo{person}{Kenny~R. Coventry}.} \bibinfo{year}{2020}\natexlab{}.
\newblock \showarticletitle{Demonstratives in Spatial Language and Social
  Interaction: An Interdisciplinary Review}.
\newblock \bibinfo{journal}{\emph{Frontiers in Psychology}}
  \bibinfo{volume}{11} (\bibinfo{year}{2020}).
\newblock
\showISSN{1664-1078}
\urldef\tempurl%
\url{https://doi.org/10.3389/fpsyg.2020.555265}
\showDOI{\tempurl}


\bibitem[DiVerdi et~al\mbox{.}(2004)]%
        {DiVerdi2004}
\bibfield{author}{\bibinfo{person}{Stephen DiVerdi}, \bibinfo{person}{Tobias
  Hollerer}, {and} \bibinfo{person}{Richard Schreyer}.}
  \bibinfo{year}{2004}\natexlab{}.
\newblock \showarticletitle{Level of Detail Interfaces}. In
  \bibinfo{booktitle}{\emph{Proceedings of the 3rd IEEE/ACM International
  Symposium on Mixed and Augmented Reality}} \emph{(\bibinfo{series}{ISMAR
  '04})}. \bibinfo{publisher}{IEEE Computer Society}, \bibinfo{address}{USA},
  \bibinfo{pages}{300–301}.
\newblock
\showISBNx{0769521916}
\urldef\tempurl%
\url{https://doi.org/10.1109/ISMAR.2004.38}
\showDOI{\tempurl}


\bibitem[Dixon(2003)]%
        {Dixon2003}
\bibfield{author}{\bibinfo{person}{R.M.W. Dixon}.}
  \bibinfo{year}{2003}\natexlab{}.
\newblock \showarticletitle{Demonstratives: A cross-linguistic typology}.
\newblock \bibinfo{journal}{\emph{Studies in Language. International Journal
  sponsored by the Foundation “Foundations of Language”}}
  \bibinfo{volume}{27}, \bibinfo{number}{1} (\bibinfo{year}{2003}),
  \bibinfo{pages}{61--112}.
\newblock
\showISSN{0378-4177}
\urldef\tempurl%
\url{https://doi.org/10.1075/sl.27.1.04dix}
\showDOI{\tempurl}


\bibitem[Drewes et~al\mbox{.}(2007)]%
        {Drewes2007}
\bibfield{author}{\bibinfo{person}{Heiko Drewes}, \bibinfo{person}{Alexander
  De~Luca}, {and} \bibinfo{person}{Albrecht Schmidt}.}
  \bibinfo{year}{2007}\natexlab{}.
\newblock \showarticletitle{Eye-Gaze Interaction for Mobile Phones}. In
  \bibinfo{booktitle}{\emph{Proceedings of the 4th International Conference on
  Mobile Technology, Applications, and Systems and the 1st International
  Symposium on Computer Human Interaction in Mobile Technology}} (Singapore)
  \emph{(\bibinfo{series}{Mobility '07})}. \bibinfo{publisher}{Association for
  Computing Machinery}, \bibinfo{address}{New York, NY, USA},
  \bibinfo{pages}{364–371}.
\newblock
\showISBNx{9781595938190}
\urldef\tempurl%
\url{https://doi.org/10.1145/1378063.1378122}
\showDOI{\tempurl}


\bibitem[Ellis et~al\mbox{.}(2011)]%
        {Ellis2011}
\bibfield{author}{\bibinfo{person}{Carolyn Ellis}, \bibinfo{person}{Tony~E.
  Adams}, {and} \bibinfo{person}{Arthur~P. Bochner}.}
  \bibinfo{year}{2011}\natexlab{}.
\newblock \showarticletitle{Autoethnography: An Overview}.
\newblock \bibinfo{journal}{\emph{Historical Social Research / Historische
  Sozialforschung}} \bibinfo{volume}{36}, \bibinfo{number}{4 (138)}
  (\bibinfo{year}{2011}), \bibinfo{pages}{273--290}.
\newblock
\showISSN{01726404}
\urldef\tempurl%
\url{http://www.jstor.org/stable/23032294}
\showURL{%
\tempurl}


\bibitem[Esteves et~al\mbox{.}(2015)]%
        {Esteves2015}
\bibfield{author}{\bibinfo{person}{Augusto Esteves}, \bibinfo{person}{Eduardo
  Velloso}, \bibinfo{person}{Andreas Bulling}, {and} \bibinfo{person}{Hans
  Gellersen}.} \bibinfo{year}{2015}\natexlab{}.
\newblock \showarticletitle{Orbits: Gaze Interaction for Smart Watches Using
  Smooth Pursuit Eye Movements}. In \bibinfo{booktitle}{\emph{Proceedings of
  the 28th Annual ACM Symposium on User Interface Software \& Technology}}
  (Charlotte, NC, USA) \emph{(\bibinfo{series}{UIST '15})}.
  \bibinfo{publisher}{Association for Computing Machinery},
  \bibinfo{address}{New York, NY, USA}, \bibinfo{pages}{457–466}.
\newblock
\showISBNx{9781450337793}
\urldef\tempurl%
\url{https://doi.org/10.1145/2807442.2807499}
\showDOI{\tempurl}


\bibitem[Findlater et~al\mbox{.}(2019)]%
        {Findlater2019}
\bibfield{author}{\bibinfo{person}{Leah Findlater}, \bibinfo{person}{Bonnie
  Chinh}, \bibinfo{person}{Dhruv Jain}, \bibinfo{person}{Jon Froehlich},
  \bibinfo{person}{Raja Kushalnagar}, {and} \bibinfo{person}{Angela~Carey
  Lin}.} \bibinfo{year}{2019}\natexlab{}.
\newblock \showarticletitle{Deaf and Hard-of-Hearing Individuals' Preferences
  for Wearable and Mobile Sound Awareness Technologies}. In
  \bibinfo{booktitle}{\emph{Proceedings of the 2019 CHI Conference on Human
  Factors in Computing Systems}} (Glasgow, Scotland Uk)
  \emph{(\bibinfo{series}{CHI '19})}. \bibinfo{publisher}{Association for
  Computing Machinery}, \bibinfo{address}{New York, NY, USA},
  \bibinfo{pages}{1–13}.
\newblock
\showISBNx{9781450359702}
\urldef\tempurl%
\url{https://doi.org/10.1145/3290605.3300276}
\showDOI{\tempurl}


\bibitem[Fox et~al\mbox{.}(2005)]%
        {Fox2005}
\bibfield{author}{\bibinfo{person}{Steve Fox}, \bibinfo{person}{Kuldeep
  Karnawat}, \bibinfo{person}{Mark Mydland}, \bibinfo{person}{Susan Dumais},
  {and} \bibinfo{person}{Thomas White}.} \bibinfo{year}{2005}\natexlab{}.
\newblock \showarticletitle{Evaluating Implicit Measures to Improve Web
  Search}.
\newblock \bibinfo{journal}{\emph{ACM Trans. Inf. Syst.}} \bibinfo{volume}{23},
  \bibinfo{number}{2} (\bibinfo{date}{apr} \bibinfo{year}{2005}),
  \bibinfo{pages}{147–168}.
\newblock
\showISSN{1046-8188}
\urldef\tempurl%
\url{https://doi.org/10.1145/1059981.1059982}
\showDOI{\tempurl}


\bibitem[Grubert et~al\mbox{.}(2017)]%
        {Grubert2017}
\bibfield{author}{\bibinfo{person}{Jens Grubert}, \bibinfo{person}{Tobias
  Langlotz}, \bibinfo{person}{Stefanie Zollmann}, {and} \bibinfo{person}{Holger
  Regenbrecht}.} \bibinfo{year}{2017}\natexlab{}.
\newblock \showarticletitle{Towards Pervasive Augmented Reality:
  Context-Awareness in Augmented Reality}.
\newblock \bibinfo{journal}{\emph{IEEE Transactions on Visualization and
  Computer Graphics}} \bibinfo{volume}{23}, \bibinfo{number}{6}
  (\bibinfo{year}{2017}), \bibinfo{pages}{1706--1724}.
\newblock
\urldef\tempurl%
\url{https://doi.org/10.1109/TVCG.2016.2543720}
\showDOI{\tempurl}


\bibitem[Guha et~al\mbox{.}(2015)]%
        {Guha2015}
\bibfield{author}{\bibinfo{person}{Ramanathan Guha}, \bibinfo{person}{Vineet
  Gupta}, \bibinfo{person}{Vivek Raghunathan}, {and}
  \bibinfo{person}{Ramakrishnan Srikant}.} \bibinfo{year}{2015}\natexlab{}.
\newblock \showarticletitle{User Modeling for a Personal Assistant}. In
  \bibinfo{booktitle}{\emph{Proceedings of the Eighth ACM International
  Conference on Web Search and Data Mining}} (Shanghai, China)
  \emph{(\bibinfo{series}{WSDM '15})}. \bibinfo{publisher}{Association for
  Computing Machinery}, \bibinfo{address}{New York, NY, USA},
  \bibinfo{pages}{275–284}.
\newblock
\showISBNx{9781450333177}
\urldef\tempurl%
\url{https://doi.org/10.1145/2684822.2685309}
\showDOI{\tempurl}


\bibitem[Guillou(2016)]%
        {Guillou2016}
\bibfield{author}{\bibinfo{person}{Liane Guillou}.}
  \bibinfo{year}{2016}\natexlab{}.
\newblock \showarticletitle{Incorporating pronoun function into statistical
  machine translation}.
\newblock


\bibitem[Guindon et~al\mbox{.}(1987)]%
        {Guindon1987}
\bibfield{author}{\bibinfo{person}{Raymonde Guindon}, \bibinfo{person}{Kelly
  Shuldberg}, {and} \bibinfo{person}{Joyce Conner}.}
  \bibinfo{year}{1987}\natexlab{}.
\newblock \showarticletitle{Grammatical and Ungrammatical Structures in
  User-Adviser Dialogues: Evidence for Sufficiency of Restricted Languages in
  Natural Language Interfaces to Advisory Systems}. In
  \bibinfo{booktitle}{\emph{25th Annual Meeting of the Association for
  Computational Linguistics}}. \bibinfo{publisher}{Association for
  Computational Linguistics}, \bibinfo{address}{Stanford, California, USA},
  \bibinfo{pages}{41--44}.
\newblock
\urldef\tempurl%
\url{https://doi.org/10.3115/981175.981181}
\showDOI{\tempurl}


\bibitem[Hertel et~al\mbox{.}(2021)]%
        {Hertel2021}
\bibfield{author}{\bibinfo{person}{Julia Hertel}, \bibinfo{person}{Sukran
  Karaosmanoglu}, \bibinfo{person}{Susanne Schmidt}, \bibinfo{person}{Julia
  Bräker}, \bibinfo{person}{Martin Semmann}, {and} \bibinfo{person}{Frank
  Steinicke}.} \bibinfo{year}{2021}\natexlab{}.
\newblock \showarticletitle{A Taxonomy of Interaction Techniques for Immersive
  Augmented Reality based on an Iterative Literature Review}. In
  \bibinfo{booktitle}{\emph{2021 IEEE International Symposium on Mixed and
  Augmented Reality (ISMAR)}}. \bibinfo{pages}{431--440}.
\newblock
\urldef\tempurl%
\url{https://doi.org/10.1109/ISMAR52148.2021.00060}
\showDOI{\tempurl}


\bibitem[Irawati et~al\mbox{.}(2006)]%
        {Irawati2006}
\bibfield{author}{\bibinfo{person}{Sylvia Irawati}, \bibinfo{person}{Scott
  Green}, \bibinfo{person}{Mark Billinghurst}, \bibinfo{person}{Andreas
  Duenser}, {and} \bibinfo{person}{Heedong Ko}.}
  \bibinfo{year}{2006}\natexlab{}.
\newblock \showarticletitle{An Evaluation of an Augmented Reality Multimodal
  Interface Using Speech and Paddle Gestures}. In
  \bibinfo{booktitle}{\emph{Advances in Artificial Reality and
  Tele-Existence}}, \bibfield{editor}{\bibinfo{person}{Zhigeng Pan},
  \bibinfo{person}{Adrian Cheok}, \bibinfo{person}{Michael Haller},
  \bibinfo{person}{Rynson W.~H. Lau}, \bibinfo{person}{Hideo Saito}, {and}
  \bibinfo{person}{Ronghua Liang}} (Eds.). \bibinfo{publisher}{Springer Berlin
  Heidelberg}, \bibinfo{address}{Berlin, Heidelberg},
  \bibinfo{pages}{272--283}.
\newblock
\showISBNx{978-3-540-49779-0}


\bibitem[Jain et~al\mbox{.}(2018a)]%
        {Jain2018b}
\bibfield{author}{\bibinfo{person}{Dhruv Jain}, \bibinfo{person}{Bonnie Chinh},
  \bibinfo{person}{Leah Findlater}, \bibinfo{person}{Raja Kushalnagar}, {and}
  \bibinfo{person}{Jon Froehlich}.} \bibinfo{year}{2018}\natexlab{a}.
\newblock \showarticletitle{Exploring Augmented Reality Approaches to Real-Time
  Captioning: A Preliminary Autoethnographic Study}. In
  \bibinfo{booktitle}{\emph{Proceedings of the 2018 ACM Conference Companion
  Publication on Designing Interactive Systems}} (Hong Kong, China)
  \emph{(\bibinfo{series}{DIS '18 Companion})}. \bibinfo{publisher}{Association
  for Computing Machinery}, \bibinfo{address}{New York, NY, USA},
  \bibinfo{pages}{7–11}.
\newblock
\showISBNx{9781450356312}
\urldef\tempurl%
\url{https://doi.org/10.1145/3197391.3205404}
\showDOI{\tempurl}


\bibitem[Jain et~al\mbox{.}(2015)]%
        {Jain2015}
\bibfield{author}{\bibinfo{person}{Dhruv Jain}, \bibinfo{person}{Leah
  Findlater}, \bibinfo{person}{Jamie Gilkeson}, \bibinfo{person}{Benjamin
  Holland}, \bibinfo{person}{Ramani Duraiswami}, \bibinfo{person}{Dmitry
  Zotkin}, \bibinfo{person}{Christian Vogler}, {and} \bibinfo{person}{Jon~E.
  Froehlich}.} \bibinfo{year}{2015}\natexlab{}.
\newblock \showarticletitle{Head-Mounted Display Visualizations to Support
  Sound Awareness for the Deaf and Hard of Hearing}. In
  \bibinfo{booktitle}{\emph{Proceedings of the 33rd Annual ACM Conference on
  Human Factors in Computing Systems}} (Seoul, Republic of Korea)
  \emph{(\bibinfo{series}{CHI '15})}. \bibinfo{publisher}{Association for
  Computing Machinery}, \bibinfo{address}{New York, NY, USA},
  \bibinfo{pages}{241–250}.
\newblock
\showISBNx{9781450331456}
\urldef\tempurl%
\url{https://doi.org/10.1145/2702123.2702393}
\showDOI{\tempurl}


\bibitem[Jain et~al\mbox{.}(2018b)]%
        {Jain2018a}
\bibfield{author}{\bibinfo{person}{Dhruv Jain}, \bibinfo{person}{Rachel Franz},
  \bibinfo{person}{Leah Findlater}, \bibinfo{person}{Jackson Cannon},
  \bibinfo{person}{Raja Kushalnagar}, {and} \bibinfo{person}{Jon Froehlich}.}
  \bibinfo{year}{2018}\natexlab{b}.
\newblock \showarticletitle{Towards Accessible Conversations in a Mobile
  Context for People Who Are Deaf and Hard of Hearing}. In
  \bibinfo{booktitle}{\emph{Proceedings of the 20th International ACM SIGACCESS
  Conference on Computers and Accessibility}} (Galway, Ireland)
  \emph{(\bibinfo{series}{ASSETS '18})}. \bibinfo{publisher}{Association for
  Computing Machinery}, \bibinfo{address}{New York, NY, USA},
  \bibinfo{pages}{81–92}.
\newblock
\showISBNx{9781450356503}
\urldef\tempurl%
\url{https://doi.org/10.1145/3234695.3236362}
\showDOI{\tempurl}


\bibitem[Jana et~al\mbox{.}(2013)]%
        {Jana2013}
\bibfield{author}{\bibinfo{person}{Suman Jana}, \bibinfo{person}{David Molnar},
  \bibinfo{person}{Alexander Moshchuk}, \bibinfo{person}{Alan Dunn},
  \bibinfo{person}{Benjamin Livshits}, \bibinfo{person}{Helen~J. Wang}, {and}
  \bibinfo{person}{Eyal Ofek}.} \bibinfo{year}{2013}\natexlab{}.
\newblock \showarticletitle{Enabling {Fine-Grained} Permissions for Augmented
  Reality Applications with Recognizers}. In \bibinfo{booktitle}{\emph{22nd
  USENIX Security Symposium (USENIX Security 13)}}. \bibinfo{publisher}{USENIX
  Association}, \bibinfo{address}{Washington, D.C.}, \bibinfo{pages}{415--430}.
\newblock
\showISBNx{978-1-931971-03-4}
\urldef\tempurl%
\url{https://www.usenix.org/conference/usenixsecurity13/technical-sessions/presentation/jana}
\showURL{%
\tempurl}


\bibitem[Jocher et~al\mbox{.}(2023)]%
        {Jocher2023}
\bibfield{author}{\bibinfo{person}{Glenn Jocher}, \bibinfo{person}{Ayush
  Chaurasia}, {and} \bibinfo{person}{Jing Qiu}.}
  \bibinfo{year}{2023}\natexlab{}.
\newblock \bibinfo{booktitle}{\emph{{YOLOv8 by Ultralytics}}}.
\newblock
\urldef\tempurl%
\url{https://github.com/ultralytics/ultralytics}
\showURL{%
\tempurl}


\bibitem[Khan et~al\mbox{.}(2022)]%
        {Khan2022}
\bibfield{author}{\bibinfo{person}{Anam~Ahmad Khan}, \bibinfo{person}{Joshua
  Newn}, \bibinfo{person}{James Bailey}, {and} \bibinfo{person}{Eduardo
  Velloso}.} \bibinfo{year}{2022}\natexlab{}.
\newblock \showarticletitle{Integrating Gaze and Speech for Enabling Implicit
  Interactions}. In \bibinfo{booktitle}{\emph{Proceedings of the 2022 CHI
  Conference on Human Factors in Computing Systems}} (New Orleans, LA, USA)
  \emph{(\bibinfo{series}{CHI '22})}. \bibinfo{publisher}{Association for
  Computing Machinery}, \bibinfo{address}{New York, NY, USA}, Article
  \bibinfo{articleno}{349}, \bibinfo{numpages}{14}~pages.
\newblock
\showISBNx{9781450391573}
\urldef\tempurl%
\url{https://doi.org/10.1145/3491102.3502134}
\showDOI{\tempurl}


\bibitem[Kong et~al\mbox{.}(2021)]%
        {Kong2021}
\bibfield{author}{\bibinfo{person}{Andy Kong}, \bibinfo{person}{Karan Ahuja},
  \bibinfo{person}{Mayank Goel}, {and} \bibinfo{person}{Chris Harrison}.}
  \bibinfo{year}{2021}\natexlab{}.
\newblock \showarticletitle{EyeMU Interactions: Gaze + IMU Gestures on Mobile
  Devices}. In \bibinfo{booktitle}{\emph{Proceedings of the 2021 International
  Conference on Multimodal Interaction}} (Montr\'{e}al, QC, Canada)
  \emph{(\bibinfo{series}{ICMI '21})}. \bibinfo{publisher}{Association for
  Computing Machinery}, \bibinfo{address}{New York, NY, USA},
  \bibinfo{pages}{577–585}.
\newblock
\showISBNx{9781450384810}
\urldef\tempurl%
\url{https://doi.org/10.1145/3462244.3479938}
\showDOI{\tempurl}


\bibitem[Koons et~al\mbox{.}(1991)]%
        {Koons1991}
\bibfield{author}{\bibinfo{person}{David~B. Koons}, \bibinfo{person}{Carlton~J.
  Sparrell}, {and} \bibinfo{person}{Kristinn~R. Th\'{o}risson}.}
  \bibinfo{year}{1991}\natexlab{}.
\newblock \showarticletitle{Integrating Simultaneous Input from Speech, Gaze,
  and Hand Gestures}. In \bibinfo{booktitle}{\emph{Proceedings of the 1991
  International Conference on Intelligent Multimedia Interfaces}} (Anaheim, CA,
  USA) \emph{(\bibinfo{series}{IMI'91})}. \bibinfo{publisher}{AAAI Press},
  \bibinfo{pages}{257–276}.
\newblock
\showISBNx{0262631504}


\bibitem[Kyt\"{o} et~al\mbox{.}(2018)]%
        {Kyto2018}
\bibfield{author}{\bibinfo{person}{Mikko Kyt\"{o}}, \bibinfo{person}{Barrett
  Ens}, \bibinfo{person}{Thammathip Piumsomboon}, \bibinfo{person}{Gun~A. Lee},
  {and} \bibinfo{person}{Mark Billinghurst}.} \bibinfo{year}{2018}\natexlab{}.
\newblock \showarticletitle{Pinpointing: Precise Head- and Eye-Based Target
  Selection for Augmented Reality}. In \bibinfo{booktitle}{\emph{Proceedings of
  the 2018 CHI Conference on Human Factors in Computing Systems}} (Montreal QC,
  Canada) \emph{(\bibinfo{series}{CHI '18})}. \bibinfo{publisher}{Association
  for Computing Machinery}, \bibinfo{address}{New York, NY, USA},
  \bibinfo{pages}{1–14}.
\newblock
\showISBNx{9781450356206}
\urldef\tempurl%
\url{https://doi.org/10.1145/3173574.3173655}
\showDOI{\tempurl}


\bibitem[L.(1992)]%
        {Oviatt1992}
\bibfield{author}{\bibinfo{person}{OVIATT~S. L.}}
  \bibinfo{year}{1992}\natexlab{}.
\newblock \showarticletitle{Pen/voice : Complementary multimodal
  communication}.
\newblock \bibinfo{journal}{\emph{Proceedings of Speeh Tech'92}}
  (\bibinfo{year}{1992}), \bibinfo{pages}{238--241}.
\newblock
\urldef\tempurl%
\url{https://cir.nii.ac.jp/crid/1571980074518368256}
\showURL{%
\tempurl}


\bibitem[Lee et~al\mbox{.}(2021)]%
        {Lee2021}
\bibfield{author}{\bibinfo{person}{Jaewook Lee}, \bibinfo{person}{Sebastian~S.
  Rodriguez}, \bibinfo{person}{Raahul Natarrajan}, \bibinfo{person}{Jacqueline
  Chen}, \bibinfo{person}{Harsh Deep}, {and} \bibinfo{person}{Alex Kirlik}.}
  \bibinfo{year}{2021}\natexlab{}.
\newblock \showarticletitle{What’s This? A Voice and Touch Multimodal
  Approach for Ambiguity Resolution in Voice Assistants}. In
  \bibinfo{booktitle}{\emph{Proceedings of the 2021 International Conference on
  Multimodal Interaction}} (Montr\'{e}al, QC, Canada)
  \emph{(\bibinfo{series}{ICMI '21})}. \bibinfo{publisher}{Association for
  Computing Machinery}, \bibinfo{address}{New York, NY, USA},
  \bibinfo{pages}{512–520}.
\newblock
\showISBNx{9781450384810}
\urldef\tempurl%
\url{https://doi.org/10.1145/3462244.3479902}
\showDOI{\tempurl}


\bibitem[Leech et~al\mbox{.}(2001)]%
        {Leech2001}
\bibfield{author}{\bibinfo{person}{Geoffrey Leech}, \bibinfo{person}{Paul
  Rayson}, {and} \bibinfo{person}{Andrew Wilson}.}
  \bibinfo{year}{2001}\natexlab{}.
\newblock \bibinfo{booktitle}{\emph{Word frequencies in written and spoken
  English: Based on the British National Corpus}}.
\newblock \bibinfo{publisher}{Routledge}.
\newblock


\bibitem[Li et~al\mbox{.}(2023)]%
        {Li2023}
\bibfield{author}{\bibinfo{person}{Junnan Li}, \bibinfo{person}{Dongxu Li},
  \bibinfo{person}{Silvio Savarese}, {and} \bibinfo{person}{Steven Hoi}.}
  \bibinfo{year}{2023}\natexlab{}.
\newblock \bibinfo{title}{BLIP-2: Bootstrapping Language-Image Pre-training
  with Frozen Image Encoders and Large Language Models}.
\newblock
\newblock
\showeprint[arxiv]{2301.12597}~[cs.CV]


\bibitem[Liao et~al\mbox{.}(2022)]%
        {Liao2022}
\bibfield{author}{\bibinfo{person}{Jian Liao}, \bibinfo{person}{Adnan Karim},
  \bibinfo{person}{Shivesh~Singh Jadon}, \bibinfo{person}{Rubaiat~Habib Kazi},
  {and} \bibinfo{person}{Ryo Suzuki}.} \bibinfo{year}{2022}\natexlab{}.
\newblock \showarticletitle{RealityTalk: Real-Time Speech-Driven Augmented
  Presentation for AR Live Storytelling}. In
  \bibinfo{booktitle}{\emph{Proceedings of the 35th Annual ACM Symposium on
  User Interface Software and Technology}} (Bend, OR, USA)
  \emph{(\bibinfo{series}{UIST '22})}. \bibinfo{publisher}{Association for
  Computing Machinery}, \bibinfo{address}{New York, NY, USA}, Article
  \bibinfo{articleno}{17}, \bibinfo{numpages}{12}~pages.
\newblock
\showISBNx{9781450393201}
\urldef\tempurl%
\url{https://doi.org/10.1145/3526113.3545702}
\showDOI{\tempurl}


\bibitem[Lin et~al\mbox{.}(2015)]%
        {Lin2015}
\bibfield{author}{\bibinfo{person}{Tsung-Yi Lin}, \bibinfo{person}{Michael
  Maire}, \bibinfo{person}{Serge Belongie}, \bibinfo{person}{Lubomir Bourdev},
  \bibinfo{person}{Ross Girshick}, \bibinfo{person}{James Hays},
  \bibinfo{person}{Pietro Perona}, \bibinfo{person}{Deva Ramanan},
  \bibinfo{person}{C.~Lawrence Zitnick}, {and} \bibinfo{person}{Piotr
  Dollár}.} \bibinfo{year}{2015}\natexlab{}.
\newblock \bibinfo{title}{Microsoft COCO: Common Objects in Context}.
\newblock
\newblock
\showeprint[arxiv]{1405.0312}~[cs.CV]


\bibitem[Lindlbauer et~al\mbox{.}(2019)]%
        {Lindlbauer2019}
\bibfield{author}{\bibinfo{person}{David Lindlbauer},
  \bibinfo{person}{Anna~Maria Feit}, {and} \bibinfo{person}{Otmar Hilliges}.}
  \bibinfo{year}{2019}\natexlab{}.
\newblock \showarticletitle{Context-Aware Online Adaptation of Mixed Reality
  Interfaces}. In \bibinfo{booktitle}{\emph{Proceedings of the 32nd Annual ACM
  Symposium on User Interface Software and Technology}} (New Orleans, LA, USA)
  \emph{(\bibinfo{series}{UIST '19})}. \bibinfo{publisher}{Association for
  Computing Machinery}, \bibinfo{address}{New York, NY, USA},
  \bibinfo{pages}{147–160}.
\newblock
\showISBNx{9781450368162}
\urldef\tempurl%
\url{https://doi.org/10.1145/3332165.3347945}
\showDOI{\tempurl}


\bibitem[Liu et~al\mbox{.}(2023)]%
        {Liu2023}
\bibfield{author}{\bibinfo{person}{Xingyu~"Bruce" Liu},
  \bibinfo{person}{Vladimir Kirilyuk}, \bibinfo{person}{Xiuxiu Yuan},
  \bibinfo{person}{Alex Olwal}, \bibinfo{person}{Peggy Chi},
  \bibinfo{person}{Xiang~"Anthony" Chen}, {and} \bibinfo{person}{Ruofei Du}.}
  \bibinfo{year}{2023}\natexlab{}.
\newblock \showarticletitle{Visual Captions: Augmenting Verbal Communication
  with On-the-Fly Visuals}. In \bibinfo{booktitle}{\emph{Proceedings of the
  2023 CHI Conference on Human Factors in Computing Systems}} (, Hamburg,
  Germany,) \emph{(\bibinfo{series}{CHI '23})}. \bibinfo{publisher}{Association
  for Computing Machinery}, \bibinfo{address}{New York, NY, USA}, Article
  \bibinfo{articleno}{108}, \bibinfo{numpages}{20}~pages.
\newblock
\showISBNx{9781450394215}
\urldef\tempurl%
\url{https://doi.org/10.1145/3544548.3581566}
\showDOI{\tempurl}


\bibitem[Lo{\'a}iciga et~al\mbox{.}(2017)]%
        {Loiciga2017}
\bibfield{author}{\bibinfo{person}{Sharid Lo{\'a}iciga}, \bibinfo{person}{Liane
  Guillou}, {and} \bibinfo{person}{Christian Hardmeier}.}
  \bibinfo{year}{2017}\natexlab{}.
\newblock \showarticletitle{What is it? Disambiguating the different readings
  of the pronoun ‘it’}. In \bibinfo{booktitle}{\emph{Conference on
  Empirical Methods in Natural Language Processing}}.
\newblock


\bibitem[Lu and Bowman(2021)]%
        {Lu2021}
\bibfield{author}{\bibinfo{person}{Feiyu Lu} {and} \bibinfo{person}{Doug~A.
  Bowman}.} \bibinfo{year}{2021}\natexlab{}.
\newblock \showarticletitle{Evaluating the Potential of Glanceable AR
  Interfaces for Authentic Everyday Uses}. In \bibinfo{booktitle}{\emph{2021
  IEEE Virtual Reality and 3D User Interfaces (VR)}}.
  \bibinfo{pages}{768--777}.
\newblock
\urldef\tempurl%
\url{https://doi.org/10.1109/VR50410.2021.00104}
\showDOI{\tempurl}


\bibitem[Lu et~al\mbox{.}(2020)]%
        {Lu2020}
\bibfield{author}{\bibinfo{person}{Feiyu Lu}, \bibinfo{person}{Shakiba Davari},
  \bibinfo{person}{Lee Lisle}, \bibinfo{person}{Yuan Li}, {and}
  \bibinfo{person}{Doug~A. Bowman}.} \bibinfo{year}{2020}\natexlab{}.
\newblock \showarticletitle{Glanceable AR: Evaluating Information Access
  Methods for Head-Worn Augmented Reality}. In \bibinfo{booktitle}{\emph{2020
  IEEE Conference on Virtual Reality and 3D User Interfaces (VR)}}.
  \bibinfo{pages}{930--939}.
\newblock
\urldef\tempurl%
\url{https://doi.org/10.1109/VR46266.2020.00113}
\showDOI{\tempurl}


\bibitem[Lystb\ae{}k et~al\mbox{.}(2022)]%
        {Lystbaek2022}
\bibfield{author}{\bibinfo{person}{Mathias~N. Lystb\ae{}k},
  \bibinfo{person}{Peter Rosenberg}, \bibinfo{person}{Ken Pfeuffer},
  \bibinfo{person}{Jens~Emil Gr\o{}nb\ae{}k}, {and} \bibinfo{person}{Hans
  Gellersen}.} \bibinfo{year}{2022}\natexlab{}.
\newblock \showarticletitle{Gaze-Hand Alignment: Combining Eye Gaze and Mid-Air
  Pointing for Interacting with Menus in Augmented Reality}.
\newblock \bibinfo{journal}{\emph{Proc. ACM Hum.-Comput. Interact.}}
  \bibinfo{volume}{6}, \bibinfo{number}{ETRA}, Article \bibinfo{articleno}{145}
  (\bibinfo{date}{may} \bibinfo{year}{2022}), \bibinfo{numpages}{18}~pages.
\newblock
\urldef\tempurl%
\url{https://doi.org/10.1145/3530886}
\showDOI{\tempurl}


\bibitem[Mardanbegi and Hansen(2011)]%
        {Mardanbegi2011}
\bibfield{author}{\bibinfo{person}{Diako Mardanbegi} {and}
  \bibinfo{person}{Dan~Witzner Hansen}.} \bibinfo{year}{2011}\natexlab{}.
\newblock \showarticletitle{Mobile Gaze-Based Screen Interaction in 3D
  Environments}. In \bibinfo{booktitle}{\emph{Proceedings of the 1st Conference
  on Novel Gaze-Controlled Applications}} (Karlskrona, Sweden)
  \emph{(\bibinfo{series}{NGCA '11})}. \bibinfo{publisher}{Association for
  Computing Machinery}, \bibinfo{address}{New York, NY, USA}, Article
  \bibinfo{articleno}{2}, \bibinfo{numpages}{4}~pages.
\newblock
\showISBNx{9781450306805}
\urldef\tempurl%
\url{https://doi.org/10.1145/1983302.1983304}
\showDOI{\tempurl}


\bibitem[Mauriello et~al\mbox{.}(2015)]%
        {Mauriello2015}
\bibfield{author}{\bibinfo{person}{Matthew~Louis Mauriello},
  \bibinfo{person}{Leyla Norooz}, {and} \bibinfo{person}{Jon~E. Froehlich}.}
  \bibinfo{year}{2015}\natexlab{}.
\newblock \showarticletitle{Understanding the Role of Thermography in Energy
  Auditing: Current Practices and the Potential for Automated Solutions}. In
  \bibinfo{booktitle}{\emph{Proceedings of the 33rd Annual ACM Conference on
  Human Factors in Computing Systems}} (Seoul, Republic of Korea)
  \emph{(\bibinfo{series}{CHI '15})}. \bibinfo{publisher}{Association for
  Computing Machinery}, \bibinfo{address}{New York, NY, USA},
  \bibinfo{pages}{1993–2002}.
\newblock
\showISBNx{9781450331456}
\urldef\tempurl%
\url{https://doi.org/10.1145/2702123.2702528}
\showDOI{\tempurl}


\bibitem[Mayer et~al\mbox{.}(2020)]%
        {Mayer2020}
\bibfield{author}{\bibinfo{person}{Sven Mayer}, \bibinfo{person}{Gierad Laput},
  {and} \bibinfo{person}{Chris Harrison}.} \bibinfo{year}{2020}\natexlab{}.
\newblock \showarticletitle{Enhancing Mobile Voice Assistants with WorldGaze}.
  In \bibinfo{booktitle}{\emph{Proceedings of the 2020 CHI Conference on Human
  Factors in Computing Systems}} (Honolulu, HI, USA)
  \emph{(\bibinfo{series}{CHI '20})}. \bibinfo{publisher}{Association for
  Computing Machinery}, \bibinfo{address}{New York, NY, USA},
  \bibinfo{pages}{1–10}.
\newblock
\showISBNx{9781450367080}
\urldef\tempurl%
\url{https://doi.org/10.1145/3313831.3376479}
\showDOI{\tempurl}


\bibitem[Miller et~al\mbox{.}(2017)]%
        {Miller2017}
\bibfield{author}{\bibinfo{person}{Ashley Miller}, \bibinfo{person}{Joan
  Malasig}, \bibinfo{person}{Brenda Castro}, \bibinfo{person}{Vicki~L. Hanson},
  \bibinfo{person}{Hugo Nicolau}, {and} \bibinfo{person}{Alessandra
  Brand\~{a}o}.} \bibinfo{year}{2017}\natexlab{}.
\newblock \showarticletitle{The Use of Smart Glasses for Lecture Comprehension
  by Deaf and Hard of Hearing Students}. In
  \bibinfo{booktitle}{\emph{Proceedings of the 2017 CHI Conference Extended
  Abstracts on Human Factors in Computing Systems}} (, Denver, Colorado, USA,)
  \emph{(\bibinfo{series}{CHI EA '17})}. \bibinfo{publisher}{Association for
  Computing Machinery}, \bibinfo{address}{New York, NY, USA},
  \bibinfo{pages}{1909–1915}.
\newblock
\showISBNx{9781450346566}
\urldef\tempurl%
\url{https://doi.org/10.1145/3027063.3053117}
\showDOI{\tempurl}


\bibitem[Miniotas et~al\mbox{.}(2006)]%
        {Miniotas2006}
\bibfield{author}{\bibinfo{person}{Darius Miniotas}, \bibinfo{person}{Oleg
  \v{S}pakov}, \bibinfo{person}{Ivan Tugoy}, {and} \bibinfo{person}{I.~Scott
  MacKenzie}.} \bibinfo{year}{2006}\natexlab{}.
\newblock \showarticletitle{Speech-Augmented Eye Gaze Interaction with Small
  Closely Spaced Targets}. In \bibinfo{booktitle}{\emph{Proceedings of the 2006
  Symposium on Eye Tracking Research \& Applications}} (San Diego, California)
  \emph{(\bibinfo{series}{ETRA '06})}. \bibinfo{publisher}{Association for
  Computing Machinery}, \bibinfo{address}{New York, NY, USA},
  \bibinfo{pages}{67–72}.
\newblock
\showISBNx{1595933050}
\urldef\tempurl%
\url{https://doi.org/10.1145/1117309.1117345}
\showDOI{\tempurl}


\bibitem[Neustaedter and Sengers(2012)]%
        {Neustaedter2021}
\bibfield{author}{\bibinfo{person}{Carman Neustaedter} {and}
  \bibinfo{person}{Phoebe Sengers}.} \bibinfo{year}{2012}\natexlab{}.
\newblock \showarticletitle{Autobiographical Design in HCI Research: Designing
  and Learning through Use-It-Yourself}. In
  \bibinfo{booktitle}{\emph{Proceedings of the Designing Interactive Systems
  Conference}} (Newcastle Upon Tyne, United Kingdom)
  \emph{(\bibinfo{series}{DIS '12})}. \bibinfo{publisher}{Association for
  Computing Machinery}, \bibinfo{address}{New York, NY, USA},
  \bibinfo{pages}{514–523}.
\newblock
\showISBNx{9781450312103}
\urldef\tempurl%
\url{https://doi.org/10.1145/2317956.2318034}
\showDOI{\tempurl}


\bibitem[Neßelrath et~al\mbox{.}(2016)]%
        {Neßelrath2016}
\bibfield{author}{\bibinfo{person}{Robert Neßelrath},
  \bibinfo{person}{Mohammad~Mehdi Moniri}, {and} \bibinfo{person}{Michael
  Feld}.} \bibinfo{year}{2016}\natexlab{}.
\newblock \showarticletitle{Combining Speech, Gaze, and Micro-gestures for the
  Multimodal Control of In-Car Functions}. In \bibinfo{booktitle}{\emph{2016
  12th International Conference on Intelligent Environments (IE)}}.
  \bibinfo{pages}{190--193}.
\newblock
\urldef\tempurl%
\url{https://doi.org/10.1109/IE.2016.42}
\showDOI{\tempurl}


\bibitem[Olson and Kemery(2020)]%
        {Olson2019}
\bibfield{author}{\bibinfo{person}{Christi Olson} {and} \bibinfo{person}{Kelli
  Kemery}.} \bibinfo{year}{2020}\natexlab{}.
\newblock \bibinfo{title}{2019 Voice report: Consumer adoption of voice
  technology and digital assistants}.
\newblock
\newblock
\urldef\tempurl%
\url{https://about.ads.microsoft.com/en-us/insights/2019-voice-report}
\showURL{%
\tempurl}


\bibitem[Olwal et~al\mbox{.}(2020)]%
        {Olwal2020}
\bibfield{author}{\bibinfo{person}{Alex Olwal}, \bibinfo{person}{Kevin Balke},
  \bibinfo{person}{Dmitrii Votintcev}, \bibinfo{person}{Thad Starner},
  \bibinfo{person}{Paula Conn}, \bibinfo{person}{Bonnie Chinh}, {and}
  \bibinfo{person}{Benoit Corda}.} \bibinfo{year}{2020}\natexlab{}.
\newblock \showarticletitle{Wearable Subtitles: Augmenting Spoken Communication
  with Lightweight Eyewear for All-Day Captioning}. In
  \bibinfo{booktitle}{\emph{Proceedings of the 33rd Annual ACM Symposium on
  User Interface Software and Technology}} (Virtual Event, USA)
  \emph{(\bibinfo{series}{UIST '20})}. \bibinfo{publisher}{Association for
  Computing Machinery}, \bibinfo{address}{New York, NY, USA},
  \bibinfo{pages}{1108–1120}.
\newblock
\showISBNx{9781450375146}
\urldef\tempurl%
\url{https://doi.org/10.1145/3379337.3415817}
\showDOI{\tempurl}


\bibitem[Olwal et~al\mbox{.}(2003)]%
        {Olwal2003}
\bibfield{author}{\bibinfo{person}{A. Olwal}, \bibinfo{person}{H. Benko}, {and}
  \bibinfo{person}{S. Feiner}.} \bibinfo{year}{2003}\natexlab{}.
\newblock \showarticletitle{SenseShapes: using statistical geometry for object
  selection in a multimodal augmented reality}. In
  \bibinfo{booktitle}{\emph{The Second IEEE and ACM International Symposium on
  Mixed and Augmented Reality, 2003. Proceedings.}} \bibinfo{pages}{300--301}.
\newblock
\urldef\tempurl%
\url{https://doi.org/10.1109/ISMAR.2003.1240730}
\showDOI{\tempurl}


\bibitem[OpenAI(2023a)]%
        {GPT35}
\bibfield{author}{\bibinfo{person}{OpenAI}.} \bibinfo{year}{2023}\natexlab{a}.
\newblock \bibinfo{title}{GPT-3.5}.
\newblock
\newblock
\urldef\tempurl%
\url{https://platform.openai.com/docs/models/gpt-3-5}
\showURL{%
\tempurl}


\bibitem[OpenAI(2023b)]%
        {GPT4}
\bibfield{author}{\bibinfo{person}{OpenAI}.} \bibinfo{year}{2023}\natexlab{b}.
\newblock \bibinfo{title}{GPT-4}.
\newblock
\newblock
\urldef\tempurl%
\url{https://openai.com/research/gpt-4}
\showURL{%
\tempurl}


\bibitem[OpenAI(2023c)]%
        {OpenAI}
\bibfield{author}{\bibinfo{person}{OpenAI}.} \bibinfo{year}{2023}\natexlab{c}.
\newblock \bibinfo{title}{Models}.
\newblock
\newblock
\urldef\tempurl%
\url{https://platform.openai.com/docs/models/overview}
\showURL{%
\tempurl}


\bibitem[Oviatt and Cohen(2000)]%
        {Oviatt2000}
\bibfield{author}{\bibinfo{person}{Sharon Oviatt} {and} \bibinfo{person}{Philip
  Cohen}.} \bibinfo{year}{2000}\natexlab{}.
\newblock \showarticletitle{Perceptual User Interfaces: Multimodal Interfaces
  That Process What Comes Naturally}.
\newblock \bibinfo{journal}{\emph{Commun. ACM}} \bibinfo{volume}{43},
  \bibinfo{number}{3} (\bibinfo{date}{mar} \bibinfo{year}{2000}),
  \bibinfo{pages}{45–53}.
\newblock
\showISSN{0001-0782}
\urldef\tempurl%
\url{https://doi.org/10.1145/330534.330538}
\showDOI{\tempurl}


\bibitem[Pei et~al\mbox{.}(2022)]%
        {Pei2022}
\bibfield{author}{\bibinfo{person}{Siyou Pei}, \bibinfo{person}{Alexander
  Chen}, \bibinfo{person}{Jaewook Lee}, {and} \bibinfo{person}{Yang Zhang}.}
  \bibinfo{year}{2022}\natexlab{}.
\newblock \showarticletitle{Hand Interfaces: Using Hands to Imitate Objects in
  AR/VR for Expressive Interactions}. In \bibinfo{booktitle}{\emph{Proceedings
  of the 2022 CHI Conference on Human Factors in Computing Systems}} (, New
  Orleans, LA, USA,) \emph{(\bibinfo{series}{CHI '22})}.
  \bibinfo{publisher}{Association for Computing Machinery},
  \bibinfo{address}{New York, NY, USA}, Article \bibinfo{articleno}{429},
  \bibinfo{numpages}{16}~pages.
\newblock
\showISBNx{9781450391573}
\urldef\tempurl%
\url{https://doi.org/10.1145/3491102.3501898}
\showDOI{\tempurl}


\bibitem[Peng et~al\mbox{.}(2018)]%
        {Peng2018}
\bibfield{author}{\bibinfo{person}{Yi-Hao Peng}, \bibinfo{person}{Ming-Wei
  Hsi}, \bibinfo{person}{Paul Taele}, \bibinfo{person}{Ting-Yu Lin},
  \bibinfo{person}{Po-En Lai}, \bibinfo{person}{Leon Hsu},
  \bibinfo{person}{Tzu-chuan Chen}, \bibinfo{person}{Te-Yen Wu},
  \bibinfo{person}{Yu-An Chen}, \bibinfo{person}{Hsien-Hui Tang}, {and}
  \bibinfo{person}{Mike~Y. Chen}.} \bibinfo{year}{2018}\natexlab{}.
\newblock \showarticletitle{SpeechBubbles: Enhancing Captioning Experiences for
  Deaf and Hard-of-Hearing People in Group Conversations}.
\newblock \bibinfo{journal}{\emph{Proceedings of the 2018 CHI Conference on
  Human Factors in Computing Systems}}.
\newblock
\urldef\tempurl%
\url{https://doi.org/10.1145/3173574.3173867}
\showDOI{\tempurl}


\bibitem[Peres et~al\mbox{.}(2013)]%
        {Peres2013}
\bibfield{author}{\bibinfo{person}{S.~Camille Peres}, \bibinfo{person}{Tri
  Pham}, {and} \bibinfo{person}{Ronald~G. Phillips}.}
  \bibinfo{year}{2013}\natexlab{}.
\newblock \showarticletitle{Validation of the System Usability Scale (SUS)}.
\newblock \bibinfo{journal}{\emph{Proceedings of the Human Factors and
  Ergonomics Society Annual Meeting}}  \bibinfo{volume}{57}
  (\bibinfo{year}{2013}), \bibinfo{pages}{192 -- 196}.
\newblock


\bibitem[Pfeuffer et~al\mbox{.}(2021)]%
        {Pfeuffer2021}
\bibfield{author}{\bibinfo{person}{Ken Pfeuffer}, \bibinfo{person}{Yasmeen
  Abdrabou}, \bibinfo{person}{Augusto Esteves}, \bibinfo{person}{Radiah Rivu},
  \bibinfo{person}{Yomna Abdelrahman}, \bibinfo{person}{Stefanie Meitner},
  \bibinfo{person}{Amr Saadi}, {and} \bibinfo{person}{Florian Alt}.}
  \bibinfo{year}{2021}\natexlab{}.
\newblock \showarticletitle{ARtention: A design space for gaze-adaptive user
  interfaces in augmented reality}.
\newblock \bibinfo{journal}{\emph{Computers \& Graphics}}  \bibinfo{volume}{95}
  (\bibinfo{year}{2021}), \bibinfo{pages}{1--12}.
\newblock
\showISSN{0097-8493}
\urldef\tempurl%
\url{https://doi.org/10.1016/j.cag.2021.01.001}
\showDOI{\tempurl}


\bibitem[Pfeuffer et~al\mbox{.}(2014)]%
        {Pfeuffer2014}
\bibfield{author}{\bibinfo{person}{Ken Pfeuffer}, \bibinfo{person}{Jason
  Alexander}, \bibinfo{person}{Ming~Ki Chong}, {and} \bibinfo{person}{Hans
  Gellersen}.} \bibinfo{year}{2014}\natexlab{}.
\newblock \showarticletitle{Gaze-Touch: Combining Gaze with Multi-Touch for
  Interaction on the Same Surface}. In \bibinfo{booktitle}{\emph{Proceedings of
  the 27th Annual ACM Symposium on User Interface Software and Technology}}
  (Honolulu, Hawaii, USA) \emph{(\bibinfo{series}{UIST '14})}.
  \bibinfo{publisher}{Association for Computing Machinery},
  \bibinfo{address}{New York, NY, USA}, \bibinfo{pages}{509–518}.
\newblock
\showISBNx{9781450330695}
\urldef\tempurl%
\url{https://doi.org/10.1145/2642918.2647397}
\showDOI{\tempurl}


\bibitem[Piumsomboon et~al\mbox{.}(2014)]%
        {Piumsomboon2014}
\bibfield{author}{\bibinfo{person}{Thammathip Piumsomboon},
  \bibinfo{person}{David Altimira}, \bibinfo{person}{Hyungon Kim},
  \bibinfo{person}{Adrian Clark}, \bibinfo{person}{Gun Lee}, {and}
  \bibinfo{person}{Mark Billinghurst}.} \bibinfo{year}{2014}\natexlab{}.
\newblock \showarticletitle{Grasp-Shell vs gesture-speech: A comparison of
  direct and indirect natural interaction techniques in augmented reality}. In
  \bibinfo{booktitle}{\emph{2014 IEEE International Symposium on Mixed and
  Augmented Reality (ISMAR)}}. \bibinfo{pages}{73--82}.
\newblock
\urldef\tempurl%
\url{https://doi.org/10.1109/ISMAR.2014.6948411}
\showDOI{\tempurl}


\bibitem[Pradhan et~al\mbox{.}(2018)]%
        {Pradhan2018}
\bibfield{author}{\bibinfo{person}{Alisha Pradhan}, \bibinfo{person}{Kanika
  Mehta}, {and} \bibinfo{person}{Leah Findlater}.}
  \bibinfo{year}{2018}\natexlab{}.
\newblock \showarticletitle{"Accessibility Came by Accident": Use of
  Voice-Controlled Intelligent Personal Assistants by People with
  Disabilities}. In \bibinfo{booktitle}{\emph{Proceedings of the 2018 CHI
  Conference on Human Factors in Computing Systems}} (Montreal QC, Canada)
  \emph{(\bibinfo{series}{CHI '18})}. \bibinfo{publisher}{Association for
  Computing Machinery}, \bibinfo{address}{New York, NY, USA},
  \bibinfo{pages}{1–13}.
\newblock
\showISBNx{9781450356206}
\urldef\tempurl%
\url{https://doi.org/10.1145/3173574.3174033}
\showDOI{\tempurl}


\bibitem[Reynolds and McDonell(2021)]%
        {Reynolds2021}
\bibfield{author}{\bibinfo{person}{Laria Reynolds} {and} \bibinfo{person}{Kyle
  McDonell}.} \bibinfo{year}{2021}\natexlab{}.
\newblock \showarticletitle{Prompt Programming for Large Language Models:
  Beyond the Few-Shot Paradigm}. In \bibinfo{booktitle}{\emph{Extended
  Abstracts of the 2021 CHI Conference on Human Factors in Computing Systems}}
  (Yokohama, Japan) \emph{(\bibinfo{series}{CHI EA '21})}.
  \bibinfo{publisher}{Association for Computing Machinery},
  \bibinfo{address}{New York, NY, USA}, Article \bibinfo{articleno}{314},
  \bibinfo{numpages}{7}~pages.
\newblock
\showISBNx{9781450380959}
\urldef\tempurl%
\url{https://doi.org/10.1145/3411763.3451760}
\showDOI{\tempurl}


\bibitem[Rivu et~al\mbox{.}(2020)]%
        {Rivu2020}
\bibfield{author}{\bibinfo{person}{Radiah Rivu}, \bibinfo{person}{Yasmeen
  Abdrabou}, \bibinfo{person}{Ken Pfeuffer}, \bibinfo{person}{Augusto Esteves},
  \bibinfo{person}{Stefanie Meitner}, {and} \bibinfo{person}{Florian Alt}.}
  \bibinfo{year}{2020}\natexlab{}.
\newblock \showarticletitle{StARe: Gaze-Assisted Face-to-Face Communication in
  Augmented Reality}. In \bibinfo{booktitle}{\emph{ACM Symposium on Eye
  Tracking Research and Applications}} (Stuttgart, Germany)
  \emph{(\bibinfo{series}{ETRA '20 Adjunct})}. \bibinfo{publisher}{Association
  for Computing Machinery}, \bibinfo{address}{New York, NY, USA}, Article
  \bibinfo{articleno}{14}, \bibinfo{numpages}{5}~pages.
\newblock
\showISBNx{9781450371353}
\urldef\tempurl%
\url{https://doi.org/10.1145/3379157.3388930}
\showDOI{\tempurl}


\bibitem[Rogers and Marshall(2017)]%
        {Rogers2017}
\bibfield{author}{\bibinfo{person}{Yvonne Rogers} {and} \bibinfo{person}{Paul
  Marshall}.} \bibinfo{year}{2017}\natexlab{}.
\newblock \bibinfo{booktitle}{\emph{Research in the wild} (\bibinfo{edition}{1}
  ed.)}.
\newblock \bibinfo{publisher}{Springer Cham}.
\newblock


\bibitem[Roider et~al\mbox{.}(2018)]%
        {Roider2018}
\bibfield{author}{\bibinfo{person}{Florian Roider}, \bibinfo{person}{Lars
  Reisig}, {and} \bibinfo{person}{Tom Gross}.} \bibinfo{year}{2018}\natexlab{}.
\newblock \showarticletitle{Just Look: The Benefits of Gaze-Activated Voice
  Input in the Car}. In \bibinfo{booktitle}{\emph{Adjunct Proceedings of the
  10th International Conference on Automotive User Interfaces and Interactive
  Vehicular Applications}} (Toronto, ON, Canada)
  \emph{(\bibinfo{series}{AutomotiveUI '18})}. \bibinfo{publisher}{Association
  for Computing Machinery}, \bibinfo{address}{New York, NY, USA},
  \bibinfo{pages}{210–214}.
\newblock
\showISBNx{9781450359474}
\urldef\tempurl%
\url{https://doi.org/10.1145/3239092.3265968}
\showDOI{\tempurl}


\bibitem[Romaniak et~al\mbox{.}(2020)]%
        {Romaniak2020}
\bibfield{author}{\bibinfo{person}{Yevhen Romaniak},
  \bibinfo{person}{Anastasiia Smielova}, \bibinfo{person}{Yevhenii Yakishyn},
  \bibinfo{person}{Valerii Dziubliuk}, \bibinfo{person}{Mykhailo Zlotnyk},
  {and} \bibinfo{person}{Oleksandr Viatchaninov}.}
  \bibinfo{year}{2020}\natexlab{}.
\newblock \showarticletitle{Nimble: Mobile Interface for a Visual Question
  Answering Augmented by Gestures}. In \bibinfo{booktitle}{\emph{Adjunct
  Proceedings of the 33rd Annual ACM Symposium on User Interface Software and
  Technology}} (Virtual Event, USA) \emph{(\bibinfo{series}{UIST '20
  Adjunct})}. \bibinfo{publisher}{Association for Computing Machinery},
  \bibinfo{address}{New York, NY, USA}, \bibinfo{pages}{129–131}.
\newblock
\showISBNx{9781450375153}
\urldef\tempurl%
\url{https://doi.org/10.1145/3379350.3416153}
\showDOI{\tempurl}


\bibitem[Ruiz et~al\mbox{.}(2010)]%
        {Ruiz2010}
\bibfield{author}{\bibinfo{person}{Natalie Ruiz}, \bibinfo{person}{Fang Chen},
  {and} \bibinfo{person}{Sharon Oviatt}.} \bibinfo{year}{2010}\natexlab{}.
\newblock \showarticletitle{Chapter 12 - Multimodal Input}.
\newblock In \bibinfo{booktitle}{\emph{Multimodal Signal Processing}},
  \bibfield{editor}{\bibinfo{person}{Jean-Philippe Thiran},
  \bibinfo{person}{Ferran Marqués}, {and} \bibinfo{person}{Hervé Bourlard}}
  (Eds.). \bibinfo{publisher}{Academic Press}, \bibinfo{address}{Oxford},
  \bibinfo{pages}{231--255}.
\newblock
\showISBNx{978-0-12-374825-6}
\urldef\tempurl%
\url{https://doi.org/10.1016/B978-0-12-374825-6.00010-1}
\showDOI{\tempurl}


\bibitem[Salvucci and Goldberg(2000)]%
        {Salvucci2000}
\bibfield{author}{\bibinfo{person}{Dario~D. Salvucci} {and}
  \bibinfo{person}{Joseph~H. Goldberg}.} \bibinfo{year}{2000}\natexlab{}.
\newblock \showarticletitle{Identifying Fixations and Saccades in Eye-Tracking
  Protocols}. In \bibinfo{booktitle}{\emph{Proceedings of the 2000 Symposium on
  Eye Tracking Research \& Applications}} (Palm Beach Gardens, Florida, USA)
  \emph{(\bibinfo{series}{ETRA '00})}. \bibinfo{publisher}{Association for
  Computing Machinery}, \bibinfo{address}{New York, NY, USA},
  \bibinfo{pages}{71–78}.
\newblock
\showISBNx{1581132808}
\urldef\tempurl%
\url{https://doi.org/10.1145/355017.355028}
\showDOI{\tempurl}


\bibitem[Saquib et~al\mbox{.}(2019)]%
        {Saquib2019}
\bibfield{author}{\bibinfo{person}{Nazmus Saquib},
  \bibinfo{person}{Rubaiat~Habib Kazi}, \bibinfo{person}{Li-Yi Wei}, {and}
  \bibinfo{person}{Wilmot Li}.} \bibinfo{year}{2019}\natexlab{}.
\newblock \showarticletitle{Interactive Body-Driven Graphics for Augmented
  Video Performance}. In \bibinfo{booktitle}{\emph{Proceedings of the 2019 CHI
  Conference on Human Factors in Computing Systems}} (Glasgow, Scotland Uk)
  \emph{(\bibinfo{series}{CHI '19})}. \bibinfo{publisher}{Association for
  Computing Machinery}, \bibinfo{address}{New York, NY, USA},
  \bibinfo{pages}{1–12}.
\newblock
\showISBNx{9781450359702}
\urldef\tempurl%
\url{https://doi.org/10.1145/3290605.3300852}
\showDOI{\tempurl}


\bibitem[Schipper and Brinkman(2017)]%
        {Schipper2017}
\bibfield{author}{\bibinfo{person}{Chris Schipper} {and} \bibinfo{person}{Bo
  Brinkman}.} \bibinfo{year}{2017}\natexlab{}.
\newblock \showarticletitle{Caption Placement on an Augmented Reality Head Worn
  Device}. In \bibinfo{booktitle}{\emph{Proceedings of the 19th International
  ACM SIGACCESS Conference on Computers and Accessibility}} (Baltimore,
  Maryland, USA) \emph{(\bibinfo{series}{ASSETS '17})}.
  \bibinfo{publisher}{Association for Computing Machinery},
  \bibinfo{address}{New York, NY, USA}, \bibinfo{pages}{365–366}.
\newblock
\showISBNx{9781450349260}
\urldef\tempurl%
\url{https://doi.org/10.1145/3132525.3134786}
\showDOI{\tempurl}


\bibitem[Shen et~al\mbox{.}(2023)]%
        {Shen2023}
\bibfield{author}{\bibinfo{person}{Yongliang Shen}, \bibinfo{person}{Kaitao
  Song}, \bibinfo{person}{Xu Tan}, \bibinfo{person}{Dongsheng Li},
  \bibinfo{person}{Weiming Lu}, {and} \bibinfo{person}{Yueting Zhuang}.}
  \bibinfo{year}{2023}\natexlab{}.
\newblock \bibinfo{title}{HuggingGPT: Solving AI Tasks with ChatGPT and its
  Friends in Hugging Face}.
\newblock
\newblock
\showeprint[arxiv]{2303.17580}~[cs.CL]


\bibitem[Williams et~al\mbox{.}(2020)]%
        {Williams2020}
\bibfield{author}{\bibinfo{person}{Adam~S. Williams}, \bibinfo{person}{Jason
  Garcia}, {and} \bibinfo{person}{Francisco Ortega}.}
  \bibinfo{year}{2020}\natexlab{}.
\newblock \showarticletitle{Understanding Multimodal User Gesture and Speech
  Behavior for Object Manipulation in Augmented Reality Using Elicitation}.
\newblock \bibinfo{journal}{\emph{IEEE Transactions on Visualization and
  Computer Graphics}} \bibinfo{volume}{26}, \bibinfo{number}{12}
  (\bibinfo{year}{2020}), \bibinfo{pages}{3479--3489}.
\newblock
\urldef\tempurl%
\url{https://doi.org/10.1109/TVCG.2020.3023566}
\showDOI{\tempurl}


\bibitem[Xu et~al\mbox{.}(2023)]%
        {Xu2023}
\bibfield{author}{\bibinfo{person}{Xuhai Xu}, \bibinfo{person}{Anna Yu},
  \bibinfo{person}{Tanya~R. Jonker}, \bibinfo{person}{Kashyap Todi},
  \bibinfo{person}{Feiyu Lu}, \bibinfo{person}{Xun Qian},
  \bibinfo{person}{Jo\~{a}o~Marcelo Evangelista~Belo}, \bibinfo{person}{Tianyi
  Wang}, \bibinfo{person}{Michelle Li}, \bibinfo{person}{Aran Mun},
  \bibinfo{person}{Te-Yen Wu}, \bibinfo{person}{Junxiao Shen},
  \bibinfo{person}{Ting Zhang}, \bibinfo{person}{Narine Kokhlikyan},
  \bibinfo{person}{Fulton Wang}, \bibinfo{person}{Paul Sorenson},
  \bibinfo{person}{Sophie Kim}, {and} \bibinfo{person}{Hrvoje Benko}.}
  \bibinfo{year}{2023}\natexlab{}.
\newblock \showarticletitle{XAIR: A Framework of Explainable AI in Augmented
  Reality}. In \bibinfo{booktitle}{\emph{Proceedings of the 2023 CHI Conference
  on Human Factors in Computing Systems}} (Hamburg, Germany)
  \emph{(\bibinfo{series}{CHI '23})}. \bibinfo{publisher}{Association for
  Computing Machinery}, \bibinfo{address}{New York, NY, USA}, Article
  \bibinfo{articleno}{202}, \bibinfo{numpages}{30}~pages.
\newblock
\showISBNx{9781450394215}
\urldef\tempurl%
\url{https://doi.org/10.1145/3544548.3581500}
\showDOI{\tempurl}


\bibitem[Yan et~al\mbox{.}(2018)]%
        {Yan2018}
\bibfield{author}{\bibinfo{person}{Yukang Yan}, \bibinfo{person}{Chun Yu},
  \bibinfo{person}{Xiaojuan Ma}, \bibinfo{person}{Xin Yi}, \bibinfo{person}{Ke
  Sun}, {and} \bibinfo{person}{Yuanchun Shi}.} \bibinfo{year}{2018}\natexlab{}.
\newblock \showarticletitle{VirtualGrasp: Leveraging Experience of Interacting
  with Physical Objects to Facilitate Digital Object Retrieval}. In
  \bibinfo{booktitle}{\emph{Proceedings of the 2018 CHI Conference on Human
  Factors in Computing Systems}} (, Montreal QC, Canada,)
  \emph{(\bibinfo{series}{CHI '18})}. \bibinfo{publisher}{Association for
  Computing Machinery}, \bibinfo{address}{New York, NY, USA},
  \bibinfo{pages}{1–13}.
\newblock
\showISBNx{9781450356206}
\urldef\tempurl%
\url{https://doi.org/10.1145/3173574.3173652}
\showDOI{\tempurl}


\bibitem[Zhai et~al\mbox{.}(1999)]%
        {Zhai1999}
\bibfield{author}{\bibinfo{person}{Shumin Zhai}, \bibinfo{person}{Carlos
  Morimoto}, {and} \bibinfo{person}{Steven Ihde}.}
  \bibinfo{year}{1999}\natexlab{}.
\newblock \showarticletitle{Manual and Gaze Input Cascaded (MAGIC) Pointing}.
  In \bibinfo{booktitle}{\emph{Proceedings of the SIGCHI Conference on Human
  Factors in Computing Systems}} (Pittsburgh, Pennsylvania, USA)
  \emph{(\bibinfo{series}{CHI '99})}. \bibinfo{publisher}{Association for
  Computing Machinery}, \bibinfo{address}{New York, NY, USA},
  \bibinfo{pages}{246–253}.
\newblock
\showISBNx{0201485591}
\urldef\tempurl%
\url{https://doi.org/10.1145/302979.303053}
\showDOI{\tempurl}


\end{thebibliography}
